\documentstyle[psfig]{mn}
\begin{document}
\hsize=6truein

\renewcommand{\thefootnote}{\fnsymbol{footnote}}

\title[]{ {\it ROSAT} PSPC observations of the outer regions of the 
Perseus cluster of galaxies}

\author[]
{\parbox[]{6.in} {S. Ettori, A.C. Fabian and D.A. White\\
\footnotesize
Institute of Astronomy, Madingley Road, Cambridge CB3 0HA \\
}}                                            
% \date{open 25 Mar 1997}
\maketitle

\begin{abstract}
We present an analysis of four off-axis {\it ROSAT} PSPC observations of
the Perseus cluster of galaxies (Abell~426). 
We detect the surface brightness profile to a radius of 80 arcmin ($\sim
2.4 h_{50}^{-1}$ Mpc) from the X-ray peak. The profile is measured in
various sectors and in three different energy bands.
Firstly, a colour analysis highlights a slight variation of $N_{\rm H}$
over the region, and cool components in the core and in the eastern
sector.
We apply the $\beta$-model to the profiles from different sectors and
present a solution to the, so-called, $\beta$-problem. The residuals from
an azimuthally-averaged profile highlight extended emission both in the
East and in the West, with estimated luminosities of about 8 and 1
$\times 10^{43}$ erg s$^{-1}$, respectively. We fit several models to the
surface brightness profile, including the one obtained from the Navarro,
Frenk and White (1995) potential. 
We obtain the best fit with the gas distribution described by a power law
in the inner, cooling region and a $\beta$-model for the extended
emission. 

Through the best-fit results and the constraints from the deprojection of
the surface brightness profiles, we define the radius where the
overdensity inside the cluster is 200 times the critical value, $r_{200}$,
at $2.7 h_{50}^{-1}$ Mpc. 
Within $2.3 h_{50}^{-1}$ Mpc ($0.85 r_{200}$), the total mass in the
Perseus cluster is $1.2 \times 10^{15} M_{\odot}$ and its gas fraction
is about 30 per cent.

\end{abstract}

\begin{keywords} 
galaxies: individual: A426 -- cooling flows -- dark matter -- X-ray:
galaxies. 
\end{keywords}

\section{INTRODUCTION} 

\begin{figure}
\psfig{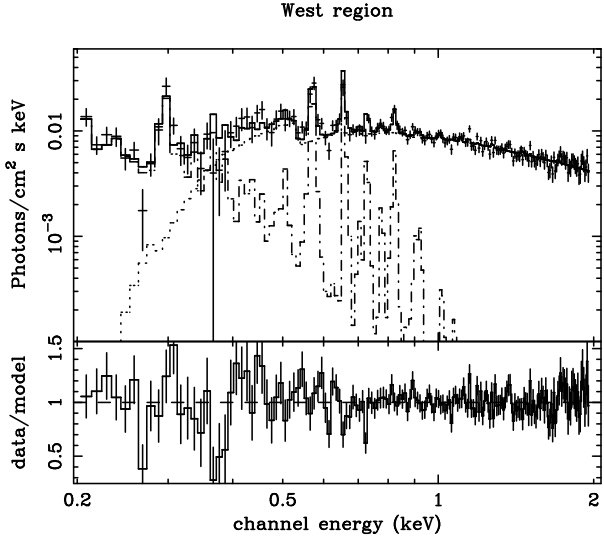}
\caption{The unfolded spectrum is shown with its various additive
components (up panel) and best-fit residuals (bottom panel).
The spectrum has been made collecting the 0.2-2 keV photons from
the central 20\arcmin--radius region in the West region.
The adopted model is the sum of a Galactic emission modelled
by a Galactic Raymond-Smith thermal plasma with Raymond-Smith
cluster emission, with metallicity of 40 per cent of the
Solar value, folded through a Galactic absorption of 12.6
$\times 10^{20}$ atoms cm$^{-2}$. The best-fit ($\chi^2$
of 211 for 173 degrees of freedom) has parameters: $kT_{\rm Gal} = 
0.22^{+0.02}_{-0.01}$ keV, $kT_{\rm cluster}=10.19^{+6.93}_{-2.51}$ keV.
} \label{fig:spec} \end{figure}   

\begin{figure}
\psfig{figure=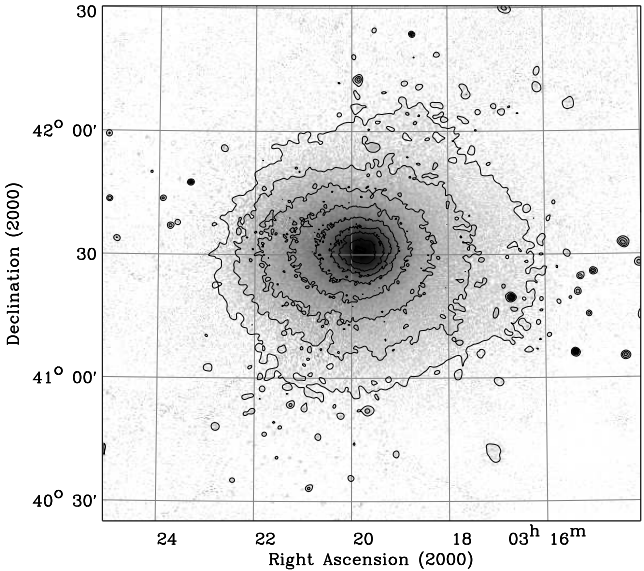,width=0.5\textwidth,angle=0}
\caption{A mosaic of the raw data of the 4 offset observations of the
Perseus Cluster, selected in the hard energy band (52--201 PI channels)
and exposure-corrected, is shown here with an adaptively
smoothed contours.
The eight contours are equally spaced in a logarithmic scale, starting
from 1.5 $\times 10^{-3}$ cts s$^{-1}$ arcmin$^{-2}$, which is 3 times the
averaged background of the four images, and increasing by a factor of 2 to
0.192 cts s$^{-1}$ arcmin$^{-2}$. 
} \label{fig:mosa} \end{figure}

The Perseus cluster (Abell~426) is a Bautz-Morgan type II--III cluster, of
richness class 2 , with a cD galaxy (NGC~1275) at its centre
($\alpha_{2000} \ 3^h 19^m 46.9^s, \delta_{2000} \ 41^{\circ} 30' 48''$). 
The cluster is located at redshift 0.0178  
(Fadda et al. 1996; 1 arcmin $\sim$ 30.1 kpc, using hereafter 
$H_0 = 50$ km s$^{-1}$ Mpc$^{-1}$, q$_0 =0.5$ and $\Lambda = 0$).

The Perseus cluster is the brightest cluster in the X-ray waveband, and so
has been extensively studied since the {\it UHURU} mission (Gursky et al.
1971, Forman et al. 1972). The emission from its intracluster medium
(ICM) is well described by a thermal gas model with a metallicity of
40 per cent of the Solar value and a temperature of 6.3 keV (cf. Allen
et al. 1992, using the {\it GINGA} satellite). The {\it Spartan 1}
detector (Snyder et al. 1990; Kowalski et al. 1993) indicates an
isothermal temperature profile, beyond the cooling flow region, 
out to 50 arcmin radius.

However, more recent work shows that the ICM has a complex morphology in
X-rays. The 505 s exposure from the {\it ROSAT} All Sky Survey
(Schwarz et al. 1992) has shown cooler emission (by a factor of 2.5--3
times lower with respect to the rest of the cluster) to the
East, corresponding to a suspected merger (recognizable
between 10 and 78 arcmin from NGC~1275). 
An analysis of the moments of the X-ray surface brightness distribution
from the {\it Einstein Observatory} IPC observations provides additional
evidence that the cluster is not yet in a relaxed state (Mohr, Fabricant
\& Geller 1993). 
Furthermore, analysis of the {\it ASCA} GIS data (Arnaud et al.
1994) indicates hotter regions to the North-West and South-West 
directions, and also a slight increases in the metal abundance.
A negative linear gradient in the metallicity has also been observed
in the {\it Spartan}~1 data (Kowalsky et al. 1993).
Furthermore, a cooling flow with a deposition rate larger than 200
$M_{\odot} \rm yr^{-1}$ is located around NGC~1275, which has an active,
radio-loud nucleus (Fabian et al. 1981, Branduardi-Raymont et al. 1981,
Primini et al. 1981, Rothschild et al. 1981).
 
In this paper, we report on the results of surveying the cluster out to
radii of 80 arcmin ($\sim 2.4$ Mpc) using a mosaic of four
off-axis, {\it ROSAT} Position Sensitive Proportional Counter (PSPC)
observations.

We present in section~2 the data and analysis of the surface
brightness profiles using both a deprojection technique and a detailed
colour analysis.
We also attempt to model the gas profile with a $\beta$-model, and 
discuss the different behaviour according to the regions and energy
bands considered.

Our purpose is to place strict constraints on the non-luminous matter in
the cluster, under the assumption of hydrostatic equilibrium, as discussed
in section 3. The main conclusions are summarized in section 4.

\section{X-RAY DATA AND ANALYSIS} 

% done by  perseus_azi.pro
% 
\begin{figure} 
\psfig{figure=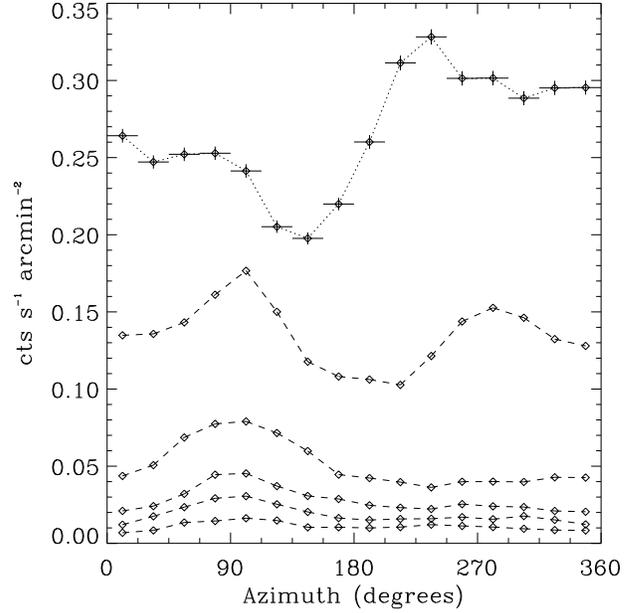,width=0.5\textwidth,angle=0}
\caption{ An azimuthal plot of the surface brightness within sectors of
22\fdg5, extracted around the X-ray peak in the on-axis image in steps of
4 arcmin for the region enclosed within $0 < r < 24$ arcmin.
For comparison the values in ordinate for the azimuthal profiles
extracted at $r>4'$ are multiplied by a factor of 3.
Due north is defined as 0\degr\, and East at 90$^{\circ}$. } 
\label{fig:azim} \end{figure}

\begin{figure}
\psfig{figure=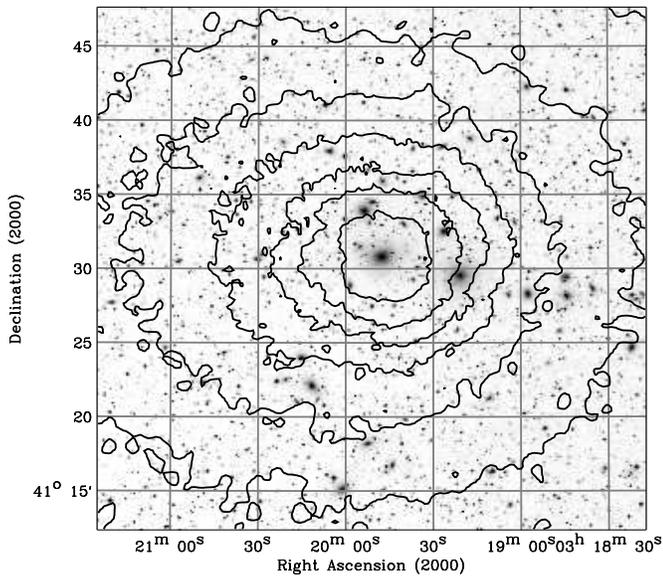,width=0.5\textwidth,angle=0}
\caption{ The adaptively smoothed contours, with the same levels as in
Fig.~\ref{fig:mosa}, are here plotted on the {\it Digitized Sky
Survey} image of the central (30 arcmin)$^2$ area of the Perseus cluster. 
The chain of bright galaxies is evident westwards, in a direction of
NGC~1272, located at about 5 arcmin from the cD, NGC~1275.
}\label{fig:opt} \end{figure}

\begin{figure*} \hspace*{-0.5cm}
\psfig{figure=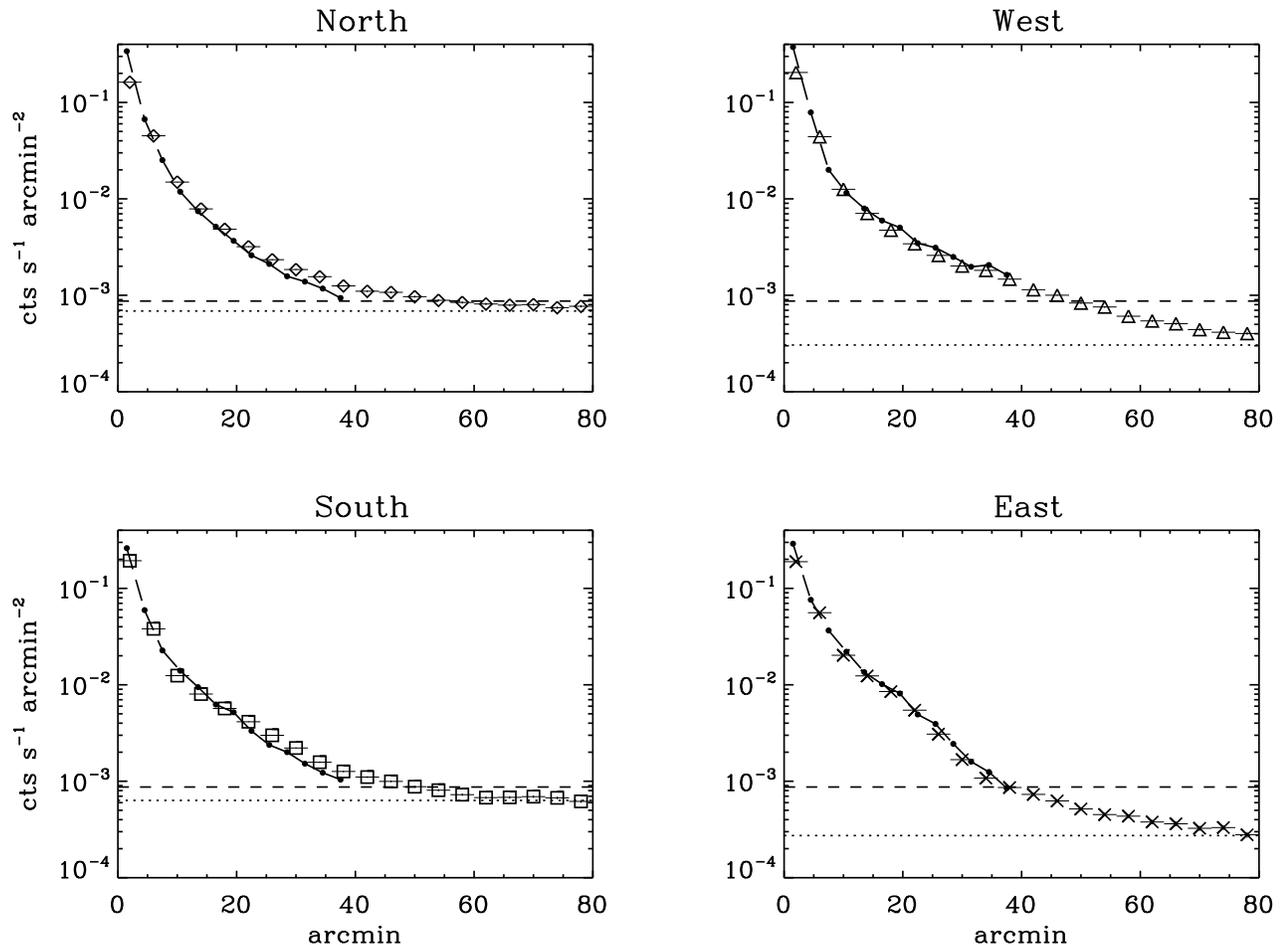,width=\textwidth,angle=90}
\caption{The surface brightness profiles in bins of 4 arcmin ware
extracted from the corresponding off-axis image and exposure-corrected
for the counts in the 0.5-2.0 keV band. 
The dotted line indicates the background value as estimated in the
45$^{\circ}$-sector opposite to the peak of the X-ray emission.
The over-plotted points are the surface brightness profiles extracted,
in bins of width of 3 arcmin, from the respective sectors of the on-axis
image. In this case, the background (dashed line) has been estimated 40
arcmin from the X-ray peak in emission. The statistical error in each bin
of the surface brightness profile is smaller than the dimension of the
point symbol adopted. 
%% The effect of the off-axis Point-Spread-Function is evident in
%% the first bin, where the on-axis profile steepen.
} \label{fig:sbkg}
\end{figure*}
 
\begin{figure*} \hspace*{1.5cm}
\psfig{figure=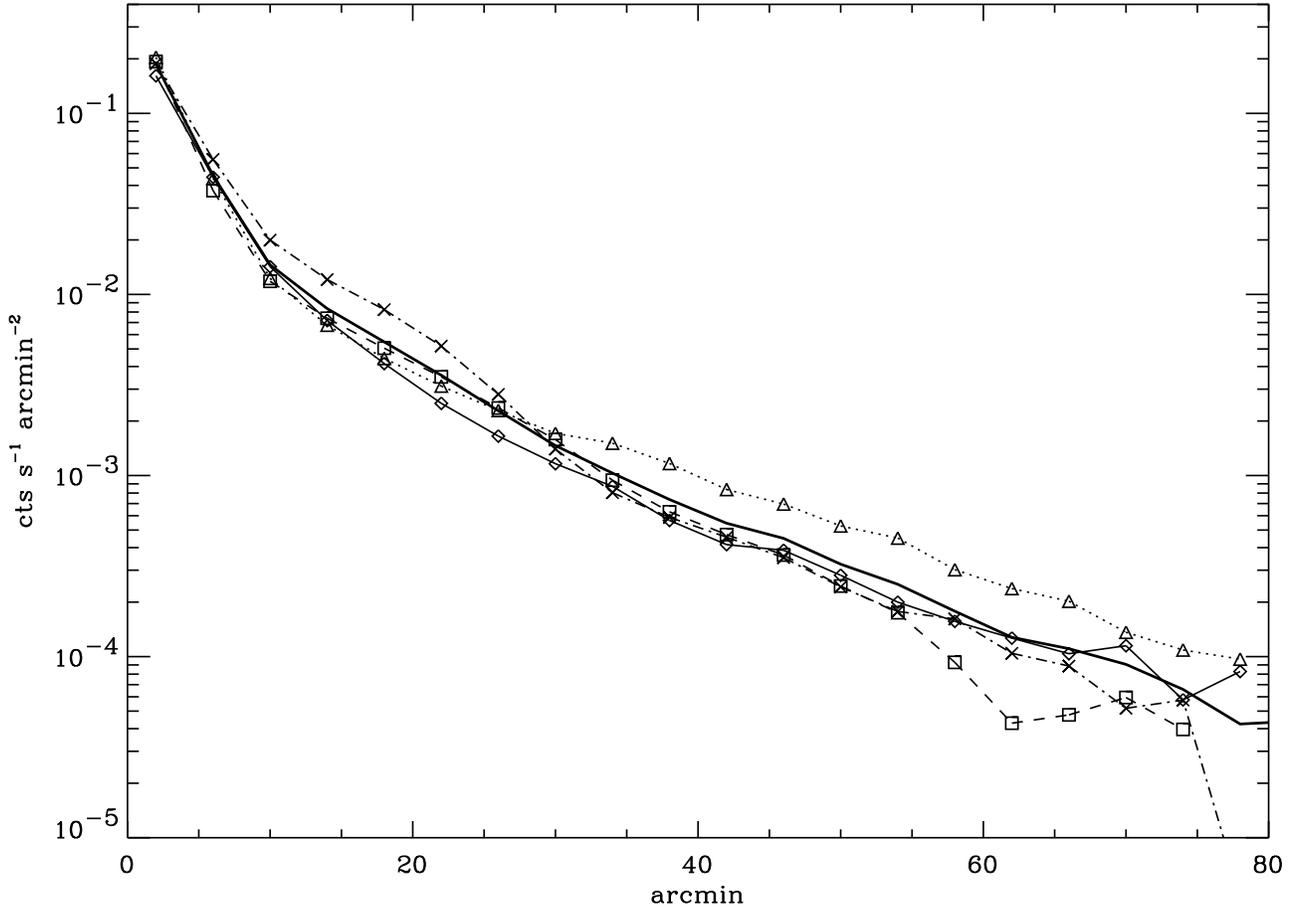,width=\textwidth,angle=90}
\caption{ Here all the background-subtracted off-axis surface brightness
profiles, selected in the 0.5-2.0 keV band, 
are plotted together (using the corresponding symbol from
Fig.~\ref{fig:sbkg})
with their average ({\it Centre-off} profile; thick solid line).
% and the azimuthally-averaged profile of the on-axis image ({\it
% Centre-on}; thick solid line that ends at 40 arcmin). 
} \label{fig:all} \end{figure*}
 
\begin{figure}
\psfig{figure=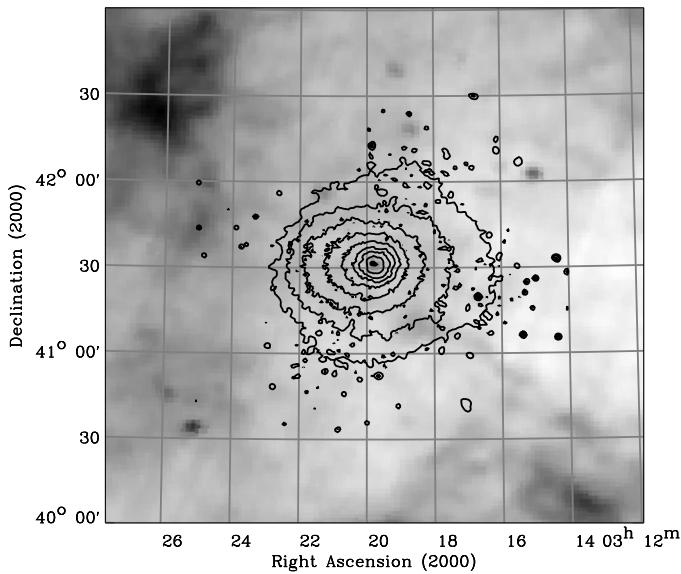,width=0.5\textwidth,angle=0} 
\caption{The eight contours of Fig.~\ref{fig:mosa} are plotted on the
3\degr $\times$ 3\degr {\it IRAS} 100 micron map. The IRAS data are 
representative of the Galactic absorption and show, on the left, the plane
of the Galaxy at latitude $b \sim -12^{\circ}$. The (minimum, maximum)
values in the {\it IRAS} map are ( 6.1, 19.2) $\times 10^6$ Jy
sr$^{-1}$. } \label{fig:iras}
\end{figure}

\begin{figure}
\psfig{figure=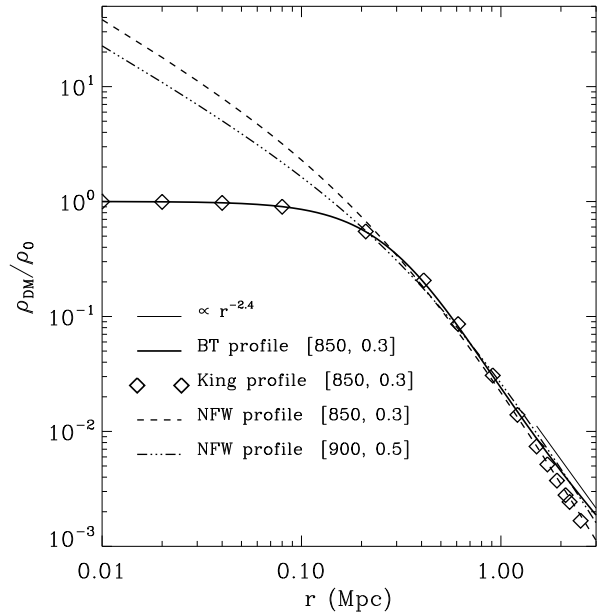,width=0.5\textwidth,angle=0}
\caption{ The Binney \& Tremaine (BT) and Navarro-Frenk-White (NFW)
dark matter profiles applied in the deprojection analysis
are here compared for different input parameters, [$\sigma$ (km s$^{-1}$), 
$r_{\rm c}$ (Mpc)]. Both of them are also compared with the King
approximation to the true isothermal sphere (King 1962).
All of them are normalized to the central value of the true isothermal
sphere profile [$\rho_0 = 9 \sigma^2 / (4 \pi G r_{\rm c}^2) = 9.05
\times 10^{-26}$ g cm$^{-3}$]. Inside the core radius, the NFW profile
does not flatter like the BT profile. 
In the outer part of the region of interest (2-3 Mpc), agreement between
the two profiles is obtained by increasing the
velocity dispersion and the core radius (or {\it scale radius}) 
in the NFW profile. Here it is also shown the dependence on $r^{-2.4}$,
as recent N-body simulations of clusters of galaxies have obtained (cf.
Thomas et al. 1998; the slope of this power law is the
mean value estimated in the range 0.1--1 $r_{180}$, where $r_{180}$ is the
radius within that the average cluster density is 180 the critical one). 
}\label{fig:dens} \end{figure}

\begin{table*}   \caption{ROSAT observation summary} 
\begin{tabular}{| l c c c c c c |} \hline
Seq. Id. & Date & Exp (s) & $\alpha_{2000}$ & $\delta_{2000}$ & Region 
& N$^{\ddagger}_{\rm H, 20}$  \\ 
\hline
\multicolumn{7}{c}{  }\\
\multicolumn{7}{c}{ ON-axis }\\
wp800186 & 1992 02 Feb &  4787 & $3^{\rm h} 19^{\rm m}$ 48\fs0 & 
$+41^{\circ} 30' 36''$ & {\it Centre} & 14.9 \\ 
\multicolumn{7}{c}{  }\\
\multicolumn{7}{c}{ OFF-axis }\\
rp800035 & 1991 18 Aug & 23669 & $3^{\rm h} 18^{\rm m}$ 28\fs7 & 
$+42^{\circ} 18' 36''$ & {\it North} & 15.2 \\
rp800032 & 1991 02 Sep & 29640 & $3^{\rm h} 15^{\rm m}$ 33\fs6 & 
$+41^{\circ} 16' 48''$ & {\it West} & 12.6 \\
rp800034 & 1991 19 Aug & 23348 & $3^{\rm h} 21^{\rm m}$ 09\fs6 & 
$+40^{\circ} 44' 24''$ & {\it South} & 14.3 \\
rp800033 & 1991 09 Feb & 14566 & $3^{\rm h} 24^{\rm m}$ 02\fs4 & 
$+41^{\circ} 46' 48''$ & {\it East} & 17.0 \\
\hline
\multicolumn{7}{l}{$^{\ddagger}$ This value (in units of 10$^{20}$ atoms
cm$^{-2}$) is the average of the HI column density,} \\ 
\multicolumn{7}{l}{calculated by interpolating the $(\alpha,
\delta)$ coordinates from the 1\degr grid maps} \\
\multicolumn{7}{l}{by Dickey \& Lockman 1990.}
\\
\end{tabular}
\end{table*}

%   ------------      TEXT    -------------------

In Table~1, we summarise the {\it ROSAT} public archive
observations that we consider in our analysis.
% with the Galactic column density value, 
% $N_{\rm H}$, measured with respect to the centre of the observation. 

The Perseus cluster centre is located at Galactic
coordinates ($l,b$: 150\fdg58, --13\fdg26). Because of its low Galactic 
latitude, Galactic emission provides a dominant contribution at very
low energy ($\sim 0.1$ keV), making (i) the corresponding energy band of
little use, and (ii) any attempt to constrain fits of the spectra
extracted from the off--axis regions difficult (for example, consider
Fig.~\ref{fig:spec}). 
% *** what happens if we fit the bkg via Rangarajan MN 277 1047 ? ***

Following Snowden et al. (1994), we correct the images by their exposure
maps in the 0.5-2.0 keV band using the Interactive
Data Language (IDL) {\it ROSAT} user-supplied libraries.
A mosaic of the four off-axis observations is shown in
Fig.~\ref{fig:mosa}, together with adaptively smoothed contours of the
exposure-corrected image.

This figure clearly shows asymmetric emission along the 
East--West axis, which is also apparent in an azimuthal plot of the
counts extracted at different radii from the central peak (cf.
Fig.~\ref{fig:azim}).
% [perseus\_azi.pro]. 
The eastern excess (around position angle, p.a.= 90\degr)
becomes significant at radii larger than 4 arcmin, whereas 
a larger contribution in emission along the West-Southwest direction
(p.a. $\sim$ 260\degr) is evident between 0 and 4 arcmin, coinciding
with the position given by Schwarz et al. (1992; cf. their Fig.~2).
The latter enhancement is also seen in a wavelet analysis of the on-axis
{\it ROSAT} PSPC image (cf. fig.~10  in Slezak, Durret \& Gerbal 1994),
and continues in the second bin (4 -- 8 arcmin). This supports the idea
of an asymmetric distribution in the gas, with a preferential orientation
along the 'chain' of bright galaxies to the West, out to a radius of 10
arcmin (cf. the optical image of the central part of the Perseus cluster 
in Fig.~\ref{fig:opt}), and with the main excess above 10 arcmin to the
East. Here, we note that further support for the idea of a merger
proceeding along East-West axis is provided by the higher intracluster gas
temperature observed by {\it ASCA} (Arnaud et al. 1994) in the North-West and
South-West regions, where the effect of the shocks are expected to
increase the cluster temperature (cf. the results of hydrodynamic
simulations in Schindler \& M\"uller 1993 and Roettinger, Loken and Burns
1997).

Moreover, an elliptical-fitting analysis with the IRAF routine {\it
ellipse} shows the same behaviour of the p.a. shown in
Fig.~\ref{fig:azim}. 
We measure an ellipticity that increases from $0.06 \pm 0.04$
at 4 arcmin to $0.22 \pm 0.07$ at 24 arcmin (beyond this radius the
error becomes larger than 50 per cent the value). The centroid 
oscillates around the X-ray peak by about 2 arcmin, first slightly to the
West, then to the East, and finally Westward again (cf.
Fig.~\ref{fig:azim}).
Apart from the Perseus cluster (cf. also Mohr et al. 1993), such
oscillations are also seen in other clusters (eg. A2256, Mohr et al. 1993;
Centaurus cluster, Allen \& Fabian 1994; A478, White et al. 1994; consider
also the studies on the behavior of the X-ray isophotes for a sample of
clusters studied by Buote \& Canizares 1996). These are indicators of
an on-going merger that causes a disturbance in the intracluster gas
\footnote[1]{Also an optical analysis shows a strong dependence of the
position of the cluster center upon limiting magnitude and galaxy
morphological types (Andreon 1994 and references therein)}.
By studying the residuals of the image, once the best-fit elliptical
isophotal model is subtracted, we find significant structure only if the
centroids of the isophotes are fixed to the position of the X-ray peak.
Allowing each isophotes centroid to vary, eliminates significant residuals.
In the following analysis, we will compare our data against the fixed
centroid model of the cluster. 

To extract the surface brightness profiles studied in the following
analysis and presented in Figs.~\ref{fig:sbkg} and \ref{fig:all},
each observation has had the region of the PSPC support ribs masked out,
as well as all detected point sources. 
Avoiding the central 1 arcmin where the non-thermal emission contributed
by NGC~1275 could be significant, the surface brightness profile has
been extracted with bins of width of 3 and 4 arcmin, for the on-axis and
the off-axis images, respectively. These bins widths allow 
smoothing over the masked circular rib at about 20 arcmin for the
on-axis images, and the Point-Spread-Function effect in the off-axis 
region of the {\it ROSAT} PSPC (cf. Fig.~5e in Hasinger et
al. 1993). 
The profiles have been extracted from four 90\degr\ sectors, North,
West, South and East, respectively, of the single on-axis image. Each
off-axis observation provides an azimuthally averaged surface brightness
profile. We have also considered an average of these profiles (called {\it
Centre-off}).
%% on-axis image ({\it Centre-on}) and the 

The background has been estimated in a 45\degr\ sector that avoids the
ribs, in a region between 40 and 45 arcmin from the centre of the image:
South-East in the on-axis image, and from the opposite side of the
field, from the peak of the cluster, in the offset images.
The X-ray peak has been estimated by applying on the correspondent image a
median filter by a square moving box with width of 5 pixels (cf. Table~2).

In Fig.~\ref{fig:sbkg} we show a comparison between the on-axis and
the off-axis profiles. Due to the limited field-of-view of the on-axis
image and the consequential difference in estimates of the
background, we only consider the off-axis
surface brightness profile in the following analysis.

%%%%%%%%%%   DBASE/center.idl 
\begin{table}   \caption{X-ray centres and their separation from NGC~1275}
\begin{tabular}{| l c c c |} \hline
Region & $\alpha_{2000}$ & $\delta_{2000}$ & $\Delta \theta$ \\
\hline
NGC~1275& $3^{\rm h} 19^{\rm m}$ 48\fs16 & +41\degr 30\arcmin 42\farcs1 &
-- \\
On-axis & $3^{\rm h} 19^{\rm m}$ 46\fs89 & +41\degr 30\arcmin 48\farcs5 &
0\fm33 \\
North   & $3^{\rm h} 19^{\rm m}$ 46\fs97 & +41\degr 32\arcmin 11\farcs8 &
1\fm52 \\
West    & $3^{\rm h} 19^{\rm m}$ 42\fs65 & +41\degr 30\arcmin 23\farcs5 &
1\fm41 \\
South   & $3^{\rm h} 19^{\rm m}$ 43\fs51 & +41\degr 29\arcmin 59\farcs1 &
1\fm37 \\
East    & $3^{\rm h} 19^{\rm m}$ 49\fs31 & +41\degr 31\arcmin 37\farcs9 &
0\fm97 \\
\hline
\end{tabular}
\end{table}

A comparison of the profiles shown in Fig.~\ref{fig:all} indicates two
enhancements with respect to the averaged value: (a) in the eastern sector
out to 25 arcmin ; (b) beyond 30 arcmin in the West. 
Galactic absorption over the region (cf. Fig.~\ref{fig:iras}) implies
a discrepancy, in the {\it ROSAT} band flux, of up to 17 per cent between
East and West, for the most and least absorbed emission respectively.
This compares with the difference in the surface brightness values of
a factor of about 2 at 1\degr radius from the X-ray centre. This
underlines that the elliptical shape of the X-ray emission is not
primarily due to the gradient in foreground absorption present over the
region.

A further discussion about the intrinsic reality of these variations is
presented in sections 2.2 and 2.3.

\subsection{The deprojection analysis}

\begin{table*}   \caption{Results of the deprojection analysis. The
$\sigma_{\rm imag}$ values come from the assumption $\beta_{\rm spec} =
\beta_{\rm imag, corr}$, where $\beta_{\rm spec}$ is the value obtained
from spectroscopy and $\beta_{\rm imag, corr}$ is the best-fit result when
a $\beta-$model is adopted, as described in subsection 2.3.2.
%%%%%%%%%%%%%%%%%%%%  beta_dpr.idl
The errors on the gas fraction, $f_{\rm gas}$, consider the uncertainty on
the temperature, $kT$, added in quadrature. 
Each column contains the two values obtained under the assumption that
the cluster gravitational potential is described by either a true
isothermal sphere (BT) or using the Navarro-Frenk-White form (NFW).} 
\begin{tabular}{| l c c c c c c c|}
\hline Region & $r_{\rm c}$ & $\sigma_{\rm dpr}$ ($\sigma_{\rm imag}$) & 
$P_{\rm out}/10^4$ &  $R_{\rm out}$ & $r_{\rm cool}$ & {\it \.{M}} & 
$f_{\rm gas}$ \\
  & Mpc & km s$^{-1}$ & K cm$^{-3}$ & Mpc & Mpc 
& $M_{\odot}$ yr$^{-1}$ & \\ \hline
 & & & & & & & \\
 & BT, NFW & BT, NFW & & & BT, NFW & BT, NFW & BT, NFW \\
 & & & & & & & \\
%% North--on  & 0.25, 0.30  & 860, 840 & 3.7 & 1.08 &
%%  0.19$^{+0.03}_{-0.06}$, 0.18$^{+0.05}_{-0.04}$  &
%%  488$^{+20}_{-19}$, 436$^{+37}_{-31}$ & 
%%  $0.206\pm 0.007$, $0.210\pm 0.006$ \\
North  & 0.25, 0.35  & 860, 860   & 1.7 & 1.93 &
  0.20$^{+0.10}_{-0.02}$, 0.20$^{+0.10}_{-0.02}$  &
  499$^{+29}_{-11}$, 474$^{+47}_{-11}$ &
  $0.31\pm0.01$, $0.31\pm 0.01$ \\
 & & ({\it 896$^{+46}_{-33}$}) & & & & & \\
 & & & & & & & \\
%% West--on & 0.10, 0.20  & 730, 700   & 6.3 & 1.08 & 
%%  0.18$^{+0.05}_{-0.04}$, 0.18$^{+0.05}_{-0.04}$ &
%%  474$^{+13}_{-16}$, 469$^{+12}_{-21}$ & 
%%  $0.378\pm0.014$, $0.338\pm0.013$  \\
West  & 0.45, 0.60  & 830, 860   & 1.6 & 2.29 & 
  0.21$^{+0.09}_{-0.03}$, 0.20$^{+0.10}_{-0.02}$ &
  554$^{+19}_{-61}$, 487$^{+5}_{-30}$ &
  $0.37\pm0.01$, $0.40\pm0.01$ \\
 & & ({\it 857$^{+26}_{-21}$}) & & & & & \\
 & & & & & & & \\
%% South--on & 0.30, 0.50  & 850, 880 & 4.5 & 1.08 & 
%%  0.17$^{+0.05}_{-0.04}$, 0.17$^{+0.06}_{-0.03}$  &
%%  387$^{+20}_{-15}$, 361$^{+29}_{-20}$  & 
%%  $0.227\pm 0.007$, $0.233\pm0.007$ \\
South  & 0.60, 1.80  & 970, 1150  & 2.1 & 1.69 &  
  0.20$^{+0.10}_{-0.02}$, 0.20$^{+0.10}_{-0.02}$  & 
  489$^{+5}_{-45}$, 452$^{+1}_{-24}$  &
  $0.21\pm 0.01$, $0.22\pm 0.01$ \\
 & & ({\it 1148$^{+96}_{-67}$}) & & & & & \\
 & & & & & & & \\
%% East--on& 0.25, 0.50  & 850, 880  & 6.3 & 0.99 & 
%%  0.18$^{+0.05}_{-0.05}$, 0.18$^{+0.04}_{-0.05}$  &
%%  463$^{+71}_{-67}$, 466$^{+77}_{-78}$ & 
%%  $0.284\pm0.005$, $0.290 \pm 0.005$\\
East  & 0.50, 0.70  & 960, 1000  & 1.7 & 1.93 &
  0.22$^{+0.09}_{-0.04}$, 0.20$^{+0.10}_{-0.02}$  &
  629$^{+14}_{-9}$, 543$^{+49}_{-13}$ &
  $0.24\pm0.01$, $0.26\pm0.01$ \\
 & & ({\it 1029$^{+54}_{-41}$}) & & & & & \\
 & & & & & & & \\
%% Centre--on &  0.25, 0.30  &  850, 800  &  5.0 & 1.08 &
%%  0.18$^{+0.05}_{-0.04}$, 0.17$^{+0.05}_{-0.04}$ & 
%%  459$^{+11}_{-9}$, 431$^{+27}_{-20}$ &
%%  $0.246 \pm 0.004$, $0.274\pm 0.005$ \\
Centre--off &  0.40, 0.70 &  890, 960 & 1.2 & 2.29 &
  0.21$^{+0.10}_{-0.02}$, 0.20$^{+0.10}_{-0.02}$ & 
  519$^{+3}_{-17}$, 477$^{+7}_{-3}$ &
  $0.30\pm 0.01$, $0.29\pm 0.01$ \\
 & & ({\it 987$^{+53}_{-41}$}) & & & & & \\
\\ \hline
\end{tabular}
\end{table*}

\begin{figure*}
\psfig{figure=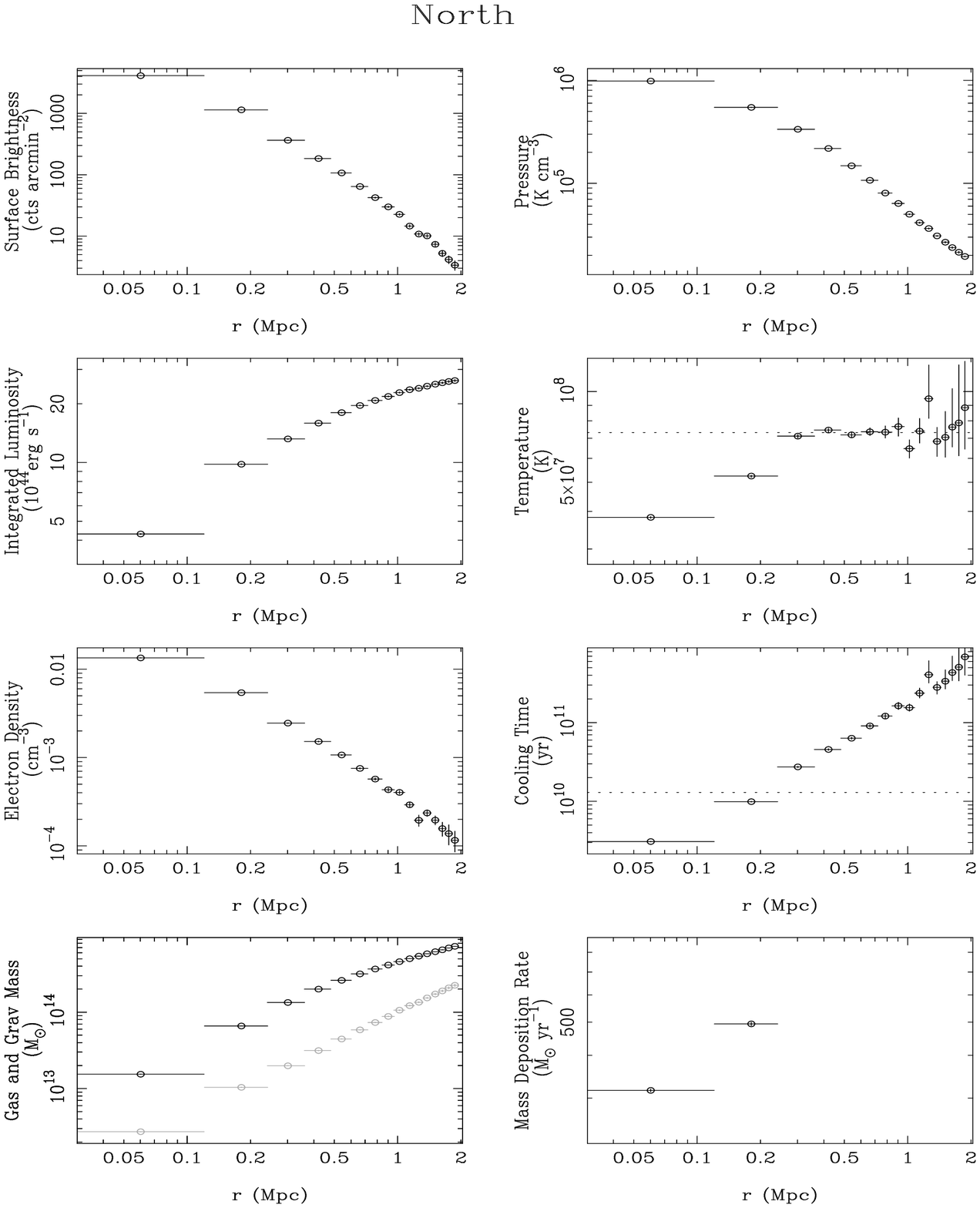,width=\textwidth,angle=0}
\caption{Plots of the deprojection results of the northward observation
of the Perseus cluster, when the potential for a true isothermal sphere is
applied.
The dotted lines in 'Temperature' and 'Cooling Time' plots represent the
adopted $T_{\rm gas}$ of 6.3 keV and the Hubble time, respectively. 
In Table 3, we list the input and output values of the deprojection
analysis.} \label{fig:dpr}
\end{figure*}
%%%%%%%%%%%%%  PS file created by /data/settori/perseus/DBASE/plot_dpr.csh
%%%%%%%%%%%%%  to format the PS file use:
%%%%%%%%%%%%%  %%BoundingBox: -18 138 650 810
%%%%%%%%%%%%%  0.077 0.062 scale
%%%%%%%%%%%%%  150 2050 translate

A deprojection analysis has been applied to the surface brightness
profiles extracted from exposure--corrected image, with photons selected
in the 0.5-2.0 keV band. (The original deprojection work by Fabian et
al. 1981 was on the Perseus cluster; for recent details on the
deprojection technique, see White et al. 1997). 
Given the assumption that the observed projected cluster emission is due
to the X-ray emitting gas in spherically symmetric shells,
the count emissivity in each radial volume shell can be calculated
analytically and compared with the predicted counts from a
bremsstrahlung emission of an optically thin gas [described by a {\it
MEKAL} model, based on the model calculations of Mewe and Kaastra
(Kaastra 1992) with Fe L calculations by Liedahl (1995)], absorbed by
intervening matter ($N_{\rm H}$ from Table~1) and convolved with the
response of the detector. 
For such an emission model, the calculated flux is proportional to the
electron density, $n_{\rm e}$, and the intracluster medium
temperature, $T_{\rm gas}$, according to the relation: $n_{\rm e}^2
f(T_{\rm gas})$ (where $f(T_{\rm gas}) \stackrel{\propto}{\sim} T_{\rm
gas}^{0.5}$, for $T_{\rm gas} \ga 3\times10^7$ K). 
Applying the perfect gas law and the equation of hydrostatic equilibrium
to the ICM, the temperature and density profiles are obtained once a
single value is known, or selected. Thus, we fix the pressure in the
outermost bin to the value that allows the resulting deprojected
temperature profile to match the observationally determined cluster
temperature of 6.3 keV. This comes from the spectral
analysis (a Raymond-Smith model plus a cooling flow component) of the data
collected from the Large Area Counter on the {\it GINGA} satellite
(Allen et al. 1992).
The gravitational potential functional form is defined by either of two
different models for the dark matter distribution: 
% (i) the King approximation to the isothermal
% sphere [\(\rho_{DM} \propto (1+x^2)^{-1.5} \) ]; 
(i) the true isothermal sphere (Binney \& Tremaine 1987); 
(ii) the Navarro-Frenk-White dark matter profile [\(\rho_{\rm DM} =
\rho_{\rm s} x^{-1}(1+x)^{-2} \); Navarro et al. 1995]. A comparison
between them for different input parameters is shown in
Fig.~\ref{fig:dens}. There, the [$\sigma, r_{\rm c}$] values for the case
of the NFW potential are the velocity dispersion of the dark matter under
the isothermal condition \footnote[3]{ $\sigma^2 = GM_{200}/(2r_{200}) =
50 H^2 r_{200}^2$, where $M_{200}, r_{200}, H$ and $G$ are the dark matter
mass at the over-density of 200, the radius at which this over-density is
reached, the Hubble constant and the Gravitational constant,
respectively}, and the {\it scale} radius, as defined in Navarro et al.
(1995). 

A gravitational contribution from the cD galaxy NGC~1275 has also been
included using a de Vaucouleurs (1948) model with a velocity dispersion
of 400 km s$^{-1}$ (from the optical $L-\sigma^4$ relation on the
brightest cluster members in Malumuth \& Kirshner 1985) and an effective
radius of 21 kpc (Schombert 1987).

In Fig.~\ref{fig:dpr}, as an example of the information obtained
by deprojecting a surface brightness profile, we show the results from
the off-axis observation of the North of the Perseus cluster.

The standard deviation confidence limits on the results summarized in
Table~3 are estimated by perturbing the surface brightness profile 100
times, according with the Poissonian error on the counts in each radial
bin. For the cooling radius, $r_{\rm cool}$, and the mass deposition
rate, {\it \.{M}}, we quote the median values, with the respective 10th
and 90th percentile limits, that come from the 100 perturbations. 

The gas mass, $M_{\rm gas}$, and the total gravitating mass, $M_{\rm
tot}$, are obtained by integrating the gas and the dark matter density,
respectively, over the volume. The gas fraction, $f_{\rm gas}$, is
defined as the ratio between the total gas mass and the gravitating mass
within the radius $R_{\rm out}$, where the surface brightness is larger
than its error by, at least, a factor of 3.

The two dark matter profiles that we adopt, the true isothermal sphere 
and the Navarro-Frenk-White profile, give good agreement in describing the
physical condition of the ICM under the hydrostatic and isothermal
assumptions, apart from a systematic reduction (about 10 per cent in the
averaged values) in the mass deposition rate when using a NFW dark matter
profile rather than the BT profile. 
% On the other hand, the total gravitating mass is lower when a NFW law is
% adopted, allowing larger gas fraction estimation by an average of 5 per
% cent.
 
However, given the low resolution that we adopt, to avoid the
technical limits on the observations (i.e., masked ribs, off-axis PSF) 
and to ensure a good signal-to-noise ratio in the outer part of
the cluster, we do not investigate further the cooling flow region, apart from
emphasizing that the results are roughly consistent with each other, and
also with recent work that indicates a cooling radius of 0.2 Mpc and {\it
\.{M}} of about 500 $M_\odot$ yr$^{-1}$ (cf. Allen \& Fabian 1997, Peres
et al. 1998).
% However, we note how the deposition rate is strictly dependent on the
% resolution adopted: when a surface brightness profile of 30\arcsec\-bins
% is used, {\it \.{M}} reduces to about 200 $M_\odot$ yr$^{-1}$.
 
%% An evident anomaly is observed in the southern sector of the
%% on-axis image. While the error-weighted mean of the three {\it
%% \.{M}} estimations that exclude the southern one is 479 $\pm$ 12
%% $M_{\odot}$ yr$^{-1}$ (when the largest error bar is considered), in
%% the southern sector we find {\it \.{M}}$\sim 390$ $M_{\odot}$
%% yr$^{-1}$, which is not consistent at about 4 times the 90\% error that
%% we adopt.
%%%%%%%%%%%%%%%%%%      err_cool.idl
%% An agreement with the other mass deposition rate estimations is
%% obtained only if one assumes that the absorption from the column
%% density, $N_{\rm H}$, is higher than the Galactic value as estimated in
%% the South area (cf. Table~1).
%% For example, using a column density of 25 $\times 10^{20}$, instead of
%% the tabulated value of 14.3 $\times 10^{20}$ atoms cm$^{-2}$, gives a
%% deposition rate of $483^{+21}_{-17}$ and a gas fraction of $0.252 \pm
%% 0.007$ for the southern on-axis region.
%% In the following subsection, this excess is tested against a colour
%% analysis of the same sector.

\subsection{The colour analysis}

\begin{figure} 
\psfig{figure=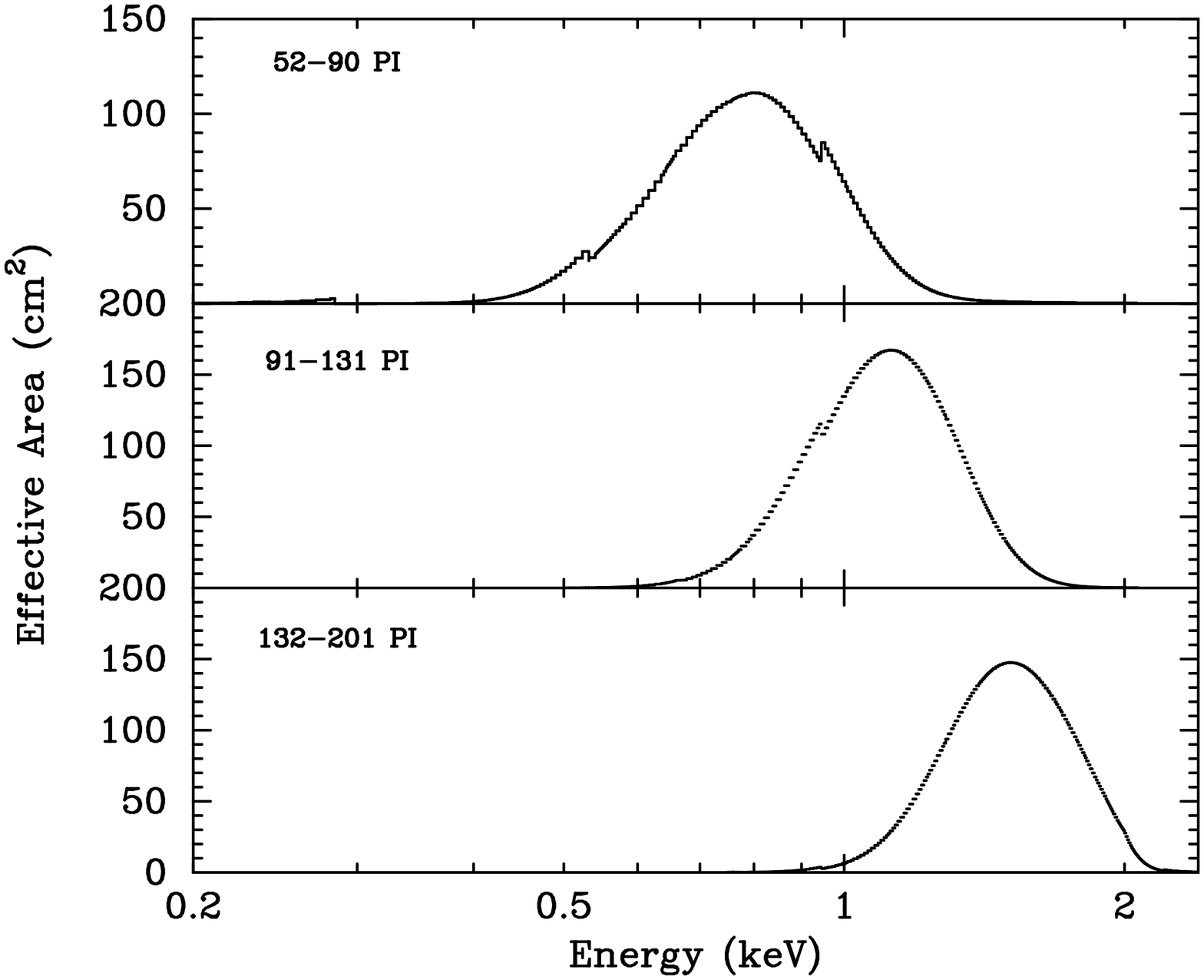,width=0.5\textwidth,angle=-90}
\caption{These are the Effective Areas of the 3 energy bands considered.
They are extracted from the Redistribution Matrix File {\tt
pspcb\_gain2\_256.rmf}. } \label{fig:effa} \end{figure}

Given the low spectral resolution of the PSPC [$\Delta E/E = 0.43 (E/0.93
{\rm keV})^{-0.5}$; {\it ROSAT} Users' Handbook 1994 \footnote[2]{also
available at the URL
heasarc.gsfc.nasa.gov/docs/rosat/ruh/handbook/handbook.html} ],
we define three independent energy bands
(or {\it colours}) where the analysis will be performed: 
(i) a soft band, which consider photons with energy enclosed between 0.52
and 0.90 keV (formed by the sum of the R4 and R5 {\it ROSAT} bands; cf.
table~1 in Snowden et al. 1994); 
(ii) a medium band, with an energy range of 0.91--1.31 keV (R6 {\it ROSAT}
band), which contains most of the Fe-L emission produced by gas at about 1
keV; (iii) a hard band (corresponding to R7) between 1.31 and 2.01 keV.
These 3 bands from on-axis PSPC response curve are shown in
Fig.~\ref{fig:effa}.

We make use of the ratios among these colours to probe the physical
properties of the Perseus cluster, once the observed ratio values are
compared with the behaviour of the same ratio in a simulated cluster
spectra. This kind of analysis using the {\it ROSAT} PSPC colours is also
comprehensively discussed in Allen \& Fabian (1997) in a study of the
cooling flows properties of a sample of clusters.

\begin{figure*} 
\psfig{figure=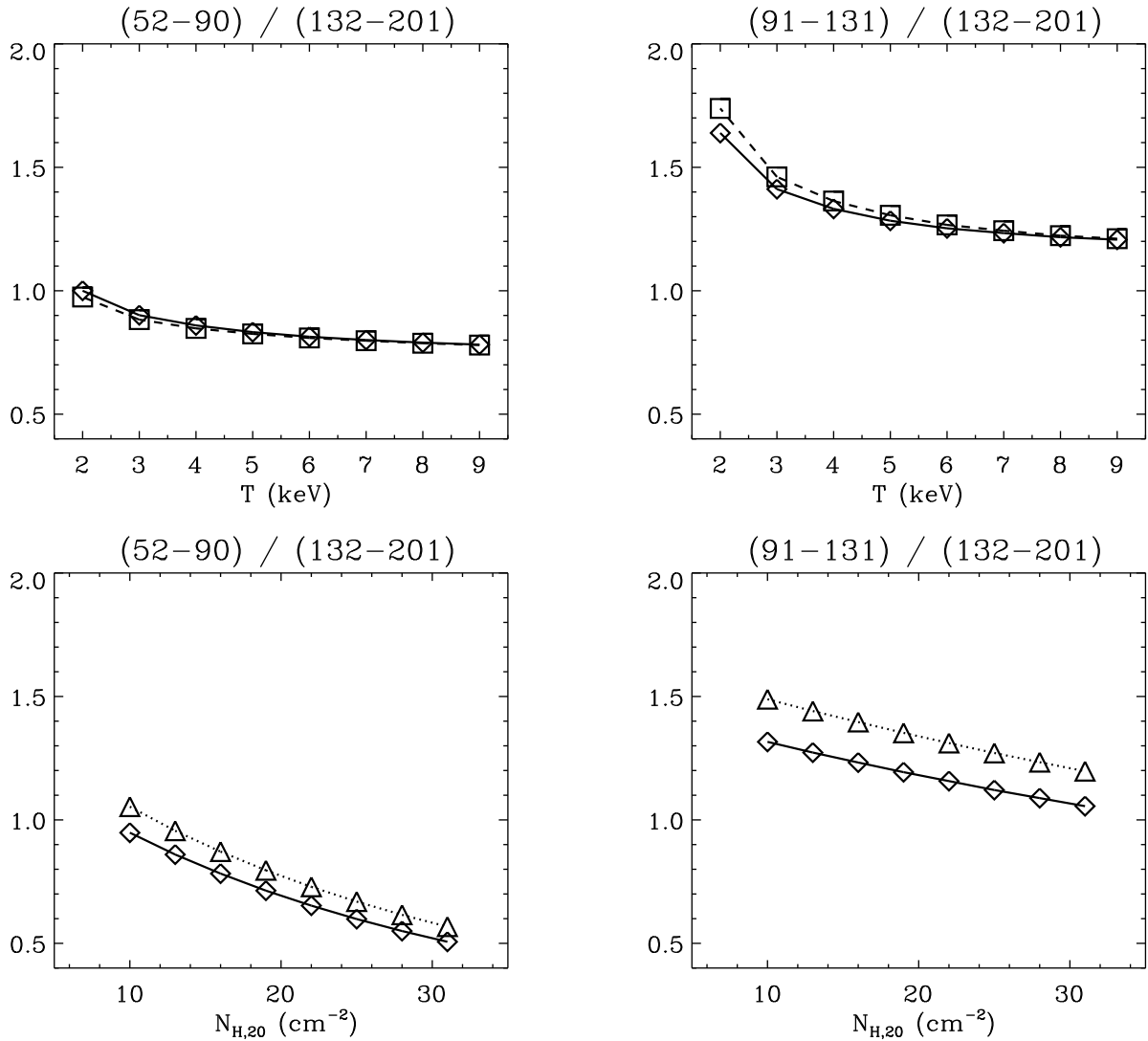,width=\textwidth,angle=0}
\caption{The {\it diamonds} connected by solid line represent the ($T =
6.3$ keV; $N_{\rm H, 20} = 14.9$; $Z/Z_{\odot} = 0.4$) case for the two
different ratios considered, when either the temperature is varied between
2 and 7 keV (upper panels) or $N_{\rm H, 20}$ is in the range 10--30
(lower panels). In the panels at the top, the {\it squares} show the
(varies; 14.9; 0.7) choice. In the panels at the bottom, the {\it triangles}
represent the (3.0; varies; 0.4) case.
These plots show that it appears difficult to detect any
gradient in metallicity [note as the squares and the diamonds
overlap]. They also show a weak dependence of the ratios upon
temperatures higher than 3 keV, and a significant
decrease when an higher column density is considered (cf. the panels
with triangles and diamonds). } \label{fig:meka}
\end{figure*}

Predicted colour ratios were calculated in XSPEC (vers. 10; Arnaud 1996)
using as reference model the plasma code of Mewe, Kaastra and Liedahl
({\it MEKAL} model; Kaastra 1992, Liedahl et al. 1995), with a
metallicity, $Z$, of either 40
or 70 per cent the Solar abundance $Z_{\odot}$, and subjected to the
photoelectric absorption provided from the model of Morrison \& McCammon
(1983). We present in Fig.~\ref{fig:meka}, for a wide range of ICM
temperature and Galactic absorption, the theoretical ratios obtained
fixing one of the parameters.

Firstly, these plots show that it appears difficult to detect any gradient
in metallicity [note that the squares and the diamonds overlap]. They
also show a weak dependence of the ratios upon temperatures higher than 3
keV. More meaningful are the steepening of the ratios at low temperatures,
particularly in the medium/hard [(91-131 PI)/(132-201 PI)] colors ratio,
and the decrease in soft/hard [(52-90 PI)/(132-201 PI)] when a higher
column density is considered (cf. the panels with triangles and diamonds).
We also note that, if $N_{\rm H}$ is assumed much lower than its {\it
true} value, this effect hides the steep increase due to the presence of a
lower gas temperature, particularly in the ratio where the softest band is
involved.

\begin{figure}
\psfig{figure=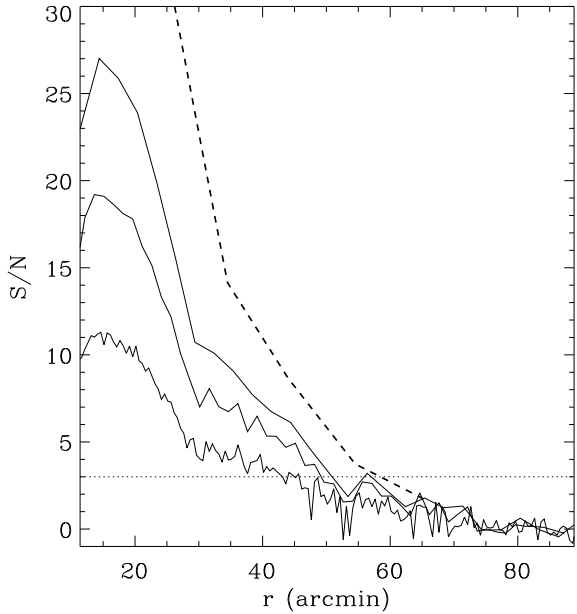,width=0.5\textwidth,angle=0}
\caption{Signal to noise ratio for Soft/Hard band ratio in the {\it East} 
with bins size of 30 arcsec, 90 arcsec and  180 arcsec (upwards solid
lines). The dashed line is for bins size of [4,5,6,...] arcmin and its
maximum is about 90 in the first 2 bins. The dotted line represents the
threshold of $S/N=3$. }
\label{fig:snr} \end{figure}

We performed a colour analysis along the four directions (North,
West, South, East) where the cluster has been observed. 
All the four offset observations allow us to extract the
surface brightness profile, $S (r)$, out to 80 arcmin from
the X-ray centre.
We use bins of [4,5,6,...] arcmin to increase the
signal-to-noise ratio at larger radii (cf. Fig.~\ref{fig:snr}).

The background was selected in a $45^\circ-$sector opposite the peak of
the X-ray emission and calculated as an average of the counts s$^{-1}$
arcmin$^{-2}$ present between 40$'$ and 45$'$. A Poissonian error is
applied to the counts of photons present in the selected region.

The errors, $\epsilon_{\rm col}$, on the colours ratio, $S_{\rm col}$, 
come from the propagation, in quadrature, of the 1--$\sigma$ uncertainty
both in each radial bin and in the background:
\begin{equation}
S_{\rm col}=\frac{S_1 - S_{{\rm bkg},1}}{S_2 - S_{{\rm bkg},2}}, \qquad
\frac{\epsilon^2_{\rm col}}{S^2_{\rm col}} = \sum_{i=1,2} \frac{\epsilon^2_i 
+ \epsilon^2_{{\rm bkg},i}}{(S_i - S_{{\rm bkg},i})^2},
\end{equation}
where $S$ and $\epsilon$ are the surface brightness value and the related
error, respectively, the index 'bkg' denotes the background value.
The sum is performed on the 2 energy bands considered. 
These errors increase towards the outer regions, becoming dominant
at radius larger than 50 arcmin (around 60 arcmin the signal to noise
ratio, $S/N$, becomes $\la 3$ for bin size of [4,5,6,...] arcmin; 
cf. Fig.~\ref{fig:snr} for the dependence of $S/N$ on the bin size.).

\begin{figure*}  \vspace*{-5cm}
\psfig{figure=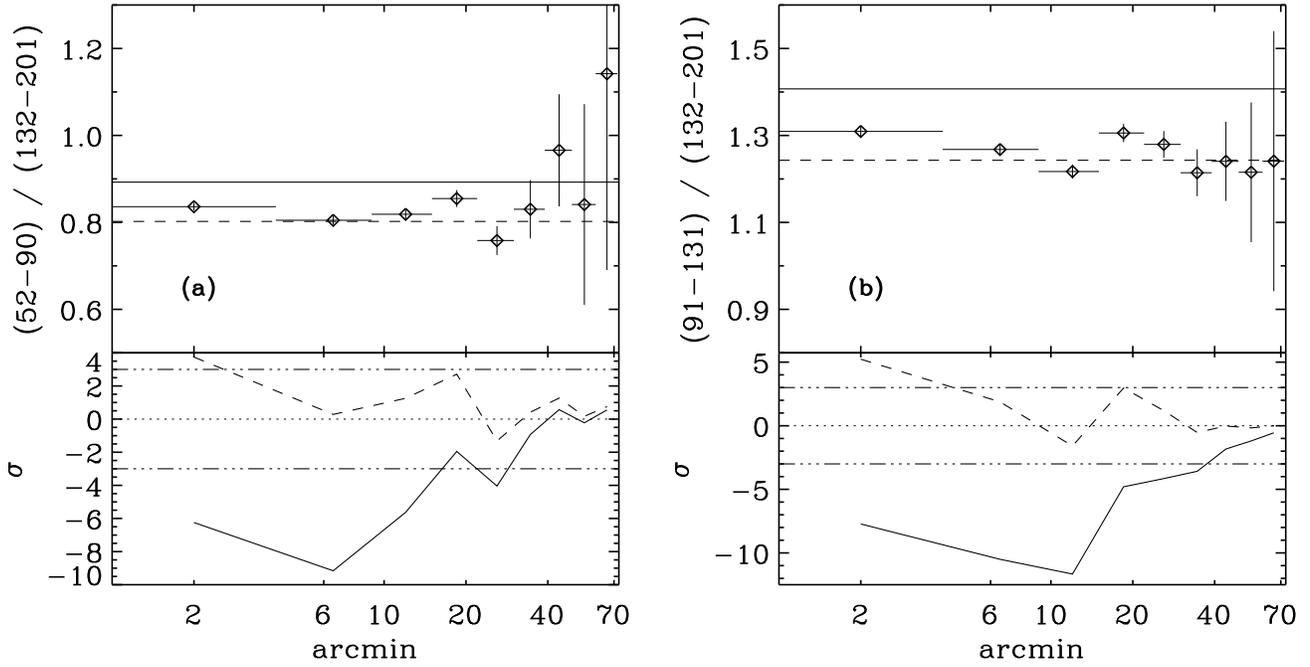,width=\textwidth,angle=90}
\caption{The colour ratios for the North region. The
Galactic absorption value used is quoted in Table~1. (Upper panels)
The colour ratios $S_{\rm col} (r)$ are compared with the predicted value
at 3 keV (solid line) and at 6.3 keV (dashed line). (Lower panels)
The deviation (in $\sigma$) from the predicted values (3 keV: solid line;
6.3 keV: dashed line) are plotted. In these panels, the dotted line shows
the zero level and the dash-dot-dot line the $\pm 3 \sigma$ level.} 
\label{fig:dev1} \end{figure*}

\begin{figure*}  \vspace*{-5cm}
\psfig{figure=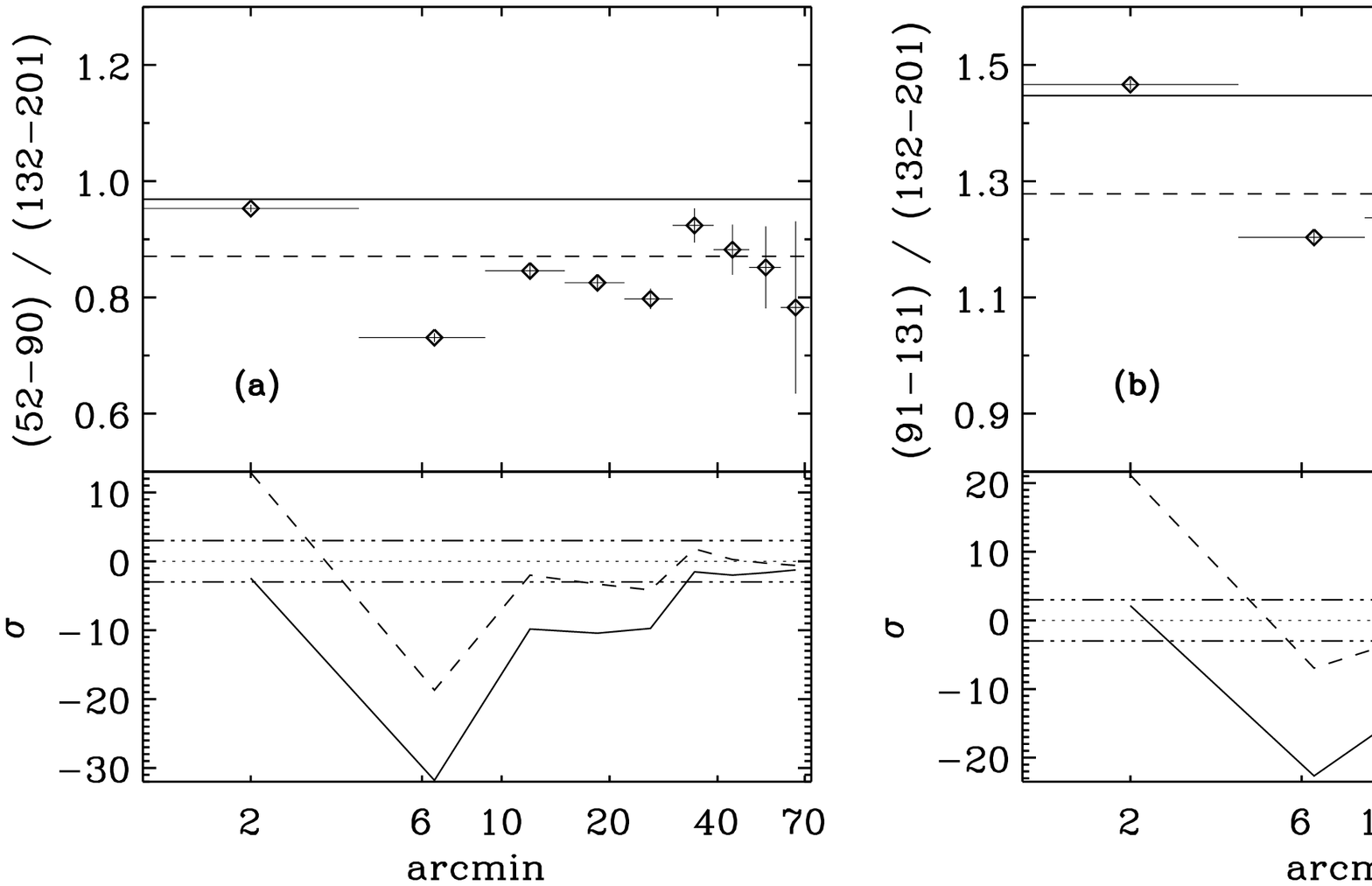,width=\textwidth,angle=90}
\caption{As Fig.~\ref{fig:dev1} for the West region. 
} \label{fig:dev2} \end{figure*}

\begin{figure*}  \vspace*{-5cm}
\psfig{figure=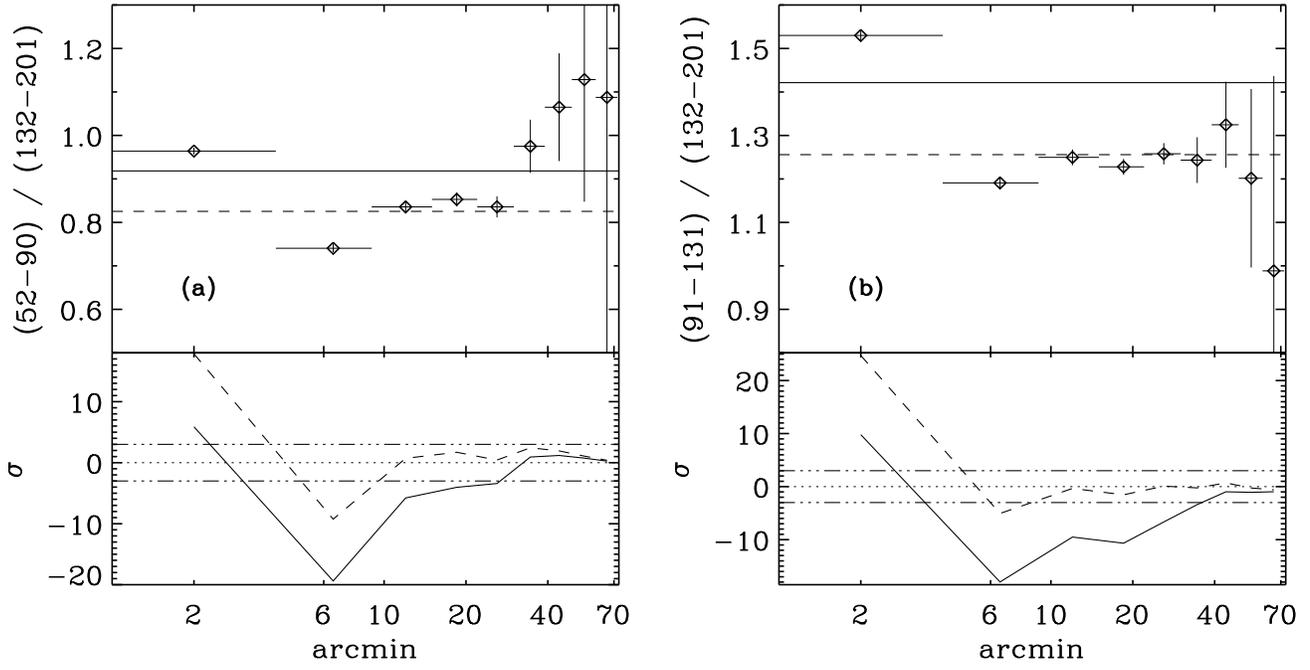,width=\textwidth,angle=90}
\caption{As Fig.~\ref{fig:dev1} for the South region. 
} \label{fig:dev3} \end{figure*}

\begin{figure*}   \vspace*{-5cm}
\psfig{figure=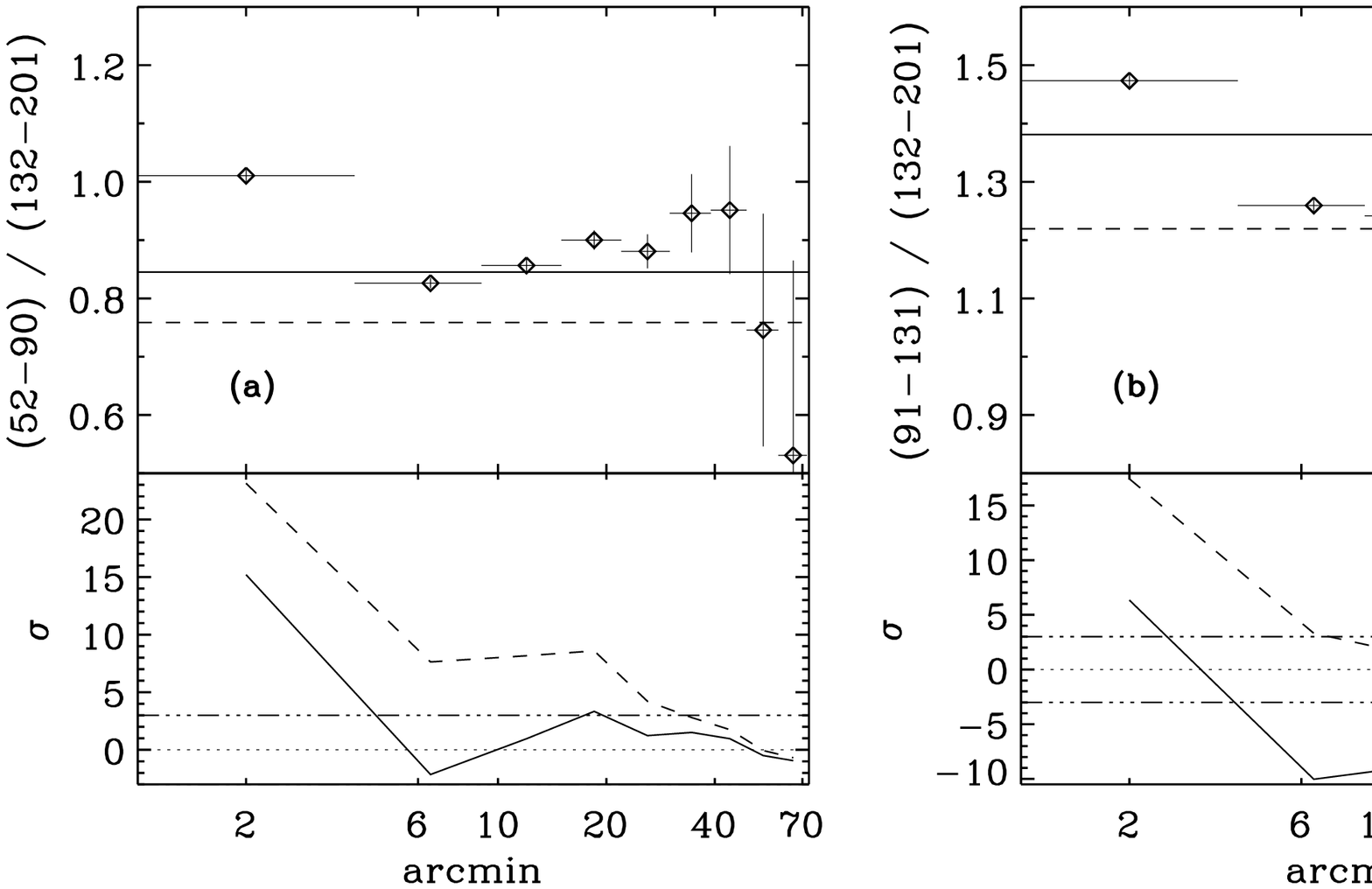,width=\textwidth,angle=90}
\caption{As Fig.~\ref{fig:dev1} for the East region. 
} \label{fig:dev4} \end{figure*}

We describe below the results of the colour analysis performed through the
ratios $S_{\rm col}(r)$. Here we note that any apparent gradient has to
be considered with respect to the error bars. We observe a slight
dependence on the  background of the colour profile in the outer region,
whilst the $\sigma$ deviations (showed in lower panels of
Fig.~\ref{fig:dev1}-\ref{fig:dev4}) are consistent whichever background
area is selected.

Our results are also compared to the conclusions from the (2.5--10
keV)/(1--2.5 keV) colour analysis on the {\it Spartan 1} data (Snyder et
al. 1990).

\begin{description}
\item {\bf North} The {\it Spartan 1} results
show in this direction a slight increase in the (2.5--10 keV)/(1--2.5 keV)
ratio moving away from the cluster centre, that they claim is related to
the observed higher Galactic absorption to the NE of Perseus (cf.
Fig.~\ref{fig:iras} ). 

In our analysis (cf. Fig.~\ref{fig:dev1}), we observe a 3$\sigma$
fluctuation in the core region due to the cooling component. Apart from
other 2$\sigma$ fluctuations around 20 arcmin, a plasma model with
$kT=6.3$ keV and $N_{\rm H}= 15.2 \times 10^{20}$ atoms cm$^{-2}$ is a
good description of the data.

\item {\bf West} In this region, {\it Spartan 1} does not show any
colour gradient.

A cool component ($\la 3$ keV) is present at high significance in the
core (panel b). The negative 3$\sigma$ deviation in the second bin can be
explained only with an increase by a factor of about 2 in the absorption
with respect to the value quoted of $N_{\rm H,20} = 12.6$ 
(Fig.~\ref{fig:dev2}). This 3$\sigma$ deviation is still evident between
15 and 30 arcmin, where the absorption remains high (cf. also
Fig.~\ref{fig:iras}) and {\it GINGA} (Arnaud et al. 1994) detects clumps
at temperatures of about 9 keV. Both of these effects decrease the colour
ratios, even if the excess in absorption is dominant and requires
roughly a $\Delta N_{\rm H,20} \sim 3$. 
Moreover, between 35 and 50 arcmin, there is a drastic increase in the
ratios due to the presence of a cool component, corresponding to the
region around the galaxy IC310.
 
%% However, the presence of a deviation less than $-3\sigma$ with respect
%% to the reference value of 3 keV puts a constraint on the lower
%% temperature limit; i.e., $kT_{\rm ICM}$ has to be greater than 3 keV
%% where we observe these large negative deviations. 

\item {\bf South} Snyder et al. (1990) observe a very flat
colour profile.

Again, the colour analysis underlines the cooling region within 4 arcmin
and an excess in absorption within 10 arcmin of about $4 \times 10^{20}$
cm$^{-2}$ with respect to the quoted value of $N_{\rm H,20} = 14.3$.
Furthermore, we obtain consistency with a $kT = 6.3$ keV above 10 arcmin,
from panel (b) (Fig.~\ref{fig:dev3}).

\item {\bf East} At radii greater than 25 arcmin , Snyder et al. (1990;
cf. their fig.~11d) observe a softening of the cluster spectrum. 

With an estimated temperature of 2.5 keV, the core presents the larger
deviation from the assumed isothermal model. Then, at least in the first 3
bins, we are overestimating $N_{\rm H}$ [from the temperature in panel
(b), we require something more close to $N_{\rm H,20} \sim 14$].
We also note that the colour ratio behaviour is more consistent with a
cooler ICM (Fig.~\ref{fig:dev4} shows less dispersion for the solid line
that represents the gas at 3 keV) between 20 and 50 arcmin. 
Here, a slight positive gradient in the colour profiles is present between
the third and sixth bin (i.e. 9--39 arcmin), underlying a
softening of the emission where the excess in surface brightness is
detected (cf. also Fig.~\ref{fig:azim} and \ref{fig:all}). 
\end{description}

To summarize, we observe variations in X-ray colour across the four
regions. There is a clear temperature decrease in the core associated
with the cooling flow. At larger radii from the centre the changes are
best explained by small variations in $N_{\rm H}$. There is no evidence
for an overall systematic temperature gradient up to 50 arcmin ($\sim 1.5$
Mpc).

\subsection{The $\beta$ problem}

\begin{table}   \caption{Best $\beta$-model fit values from previous work.
The errors are at the 90 per cent confidence level. The labels in column
{\it ref} indicate:
Cr97 (Cruddace et al. 1997), Sc92 (Schwarz et al. 1992), Sn90 (Snyder et
al. 1990), JF84 (Jones \& Forman 1984), Br81 (Branduardi-Raymont et al.
1981); Cr97 and Sn90 apply two
different method on the same data from the {\it Spartan I}: Cr97 bin the
spatial and the spectral counts distribution and fit a three-dimensional
model of the cluster structure, obtained by the Raymond-Smith code of the
plasma emissivity; Sn90 use a $\beta$-model on the surface brightness
profile. They quote a $\chi^2$ value of 287.2 (with 273
degree-of-freedom) and 20.1 (19), respectively. The
latter method is also used by Sc92 on the {\it ROSAT} All 
Sky Survey data, and by JF84 and Br81 on the {\it Einstein} IPC data (cf.
also fig.~8 in Sn90). In Sc92, both errors and $\chi^2$ values are not
quoted. In JF84, the $\chi^2$ value is not indicated. Also in Br81, no
$\chi^2$ value is indicated and the core radius is
fixed to the one determined from the optical galaxy distribution [using
a King model, Kent \& Sargent (1983) find a core radius of 11 arcmin].} 
\begin{tabular}{| l c c c |} \hline
Region & $\beta$ & $r_{\rm c}$ & ref \\ \hline
$r<50'$ & 0.73$^{+0.10}_{-0.09}$  & $13.3'^{+4.8'}_{-4.1'}$ & Cr97 \\
$10'< r <78'$ & 0.52 & $1.9'$ & Sc92 \\ 
$8'< r <50'$ & 0.72$^{+0.08}_{-0.05}$ & $12.5'^{+3.1'}_{-2.8'}$ &
Sn90 \\
$3.2'< r < 32'$ & 0.57 $\pm 0.03$ & $8.9' \pm 1.7'$ & JF84 \\
$10'<r<17'$ & 0.59 $\pm 0.02$ & $8'$ (fix) & Br81 \\ \hline
\end{tabular}
\end{table}

Cavaliere \& Fusco-Femiano (1976) pointed out that the scale height for
galaxies and gas in a cluster may be calculated from optical
velocity dispersion $\sigma_{\rm opt}$ and X-ray spectral temperature 
$kT$ as 
\begin{equation} 
\beta_{\rm spec} = \frac{\mu m_{\rm p} \sigma_{\rm opt}^2}{kT},
\end{equation}
and, for an isothermal spherically-symmetric X-ray
emission, from the surface brightness profile in the functional form (when
$r \gg r_{\rm c}$) 
\begin{equation}
S (r) = S_0 \left[1+\left(\frac{r}{r_{\rm c}} \right)^2
\right]^{0.5 -3\beta_{\rm imag}}.
\end{equation}  

Edge \& Stewart (1991) calculated from the {\it EXOSAT} data on the
Perseus cluster a $\beta_{\rm spec}$ of 1.78$^{+0.48}_{-0.34}$. Using the
more recent reference values (with corresponding 1-$\sigma$ errors) on
$\sigma_{\rm opt}$ of 1026$^{+106}_{-64}$ km s$^{-1}$ (Fadda et al.
1996) and $kT= 6.33^{+0.13}_{-0.11}$ keV  (Allen et al. 1992), we obtain
$\beta_{\rm spec}=1.09^{+0.23}_{-0.14}$. The significant decrease comes
mainly from a much more careful determination of $\sigma_{\rm opt}$, once
substructures, gaps in galaxy velocity distribution and more robust
statistical tools
are considered.
  
However, this estimate is much larger, for example, than
$\beta_{\rm imag}$ of about 0.52, obtained from the azimuthally
averaged profile of the cluster surface emission (Schwarz et al. 1992;
but also cf. Table~4). 
The disagreement between $\beta_{\rm spec}$ and $\beta_{\rm imag}$
values is the reason for the so-called $\beta$ {\it problem} in the
Perseus cluster (cf. also Allen et al. 1992). In particular, the
non-symmetric emission from the cluster (as first pointed out by
Branduardi-Raymont et al. 1981) traces a recent merger, that, together
with the excess of gas to the East and of galaxies to the West, seems to
emphasize the dependence of the $\beta-$problem on the sector analysed. In
fact, Schwarz et al. (1992) observed a value of $\beta_{\rm imag} \sim
0.5$ to West and North, and of 1.27$^{+0.29}_{-0.18}$ to East (in their
paper, there is no mention of the $\beta$ value to South).

We re-examine this issue, using our offset profiles out to about
80 arcmin from the X-ray peak. They are binned in width of [
4, 3, 2, 1, (all the remaining at) 1] arcmin, due to the PSF behaviour in
the outer part of PSPC. 

First, an attempt is made to fit the profiles over the whole range
of [ 0, 80] arcmin (cf. Table~5). Here, we note that, applying
a $\beta-$model, the goodness-of-fit is generally very poor, in a
$\chi^2$ sense. This is due to the fact that the statistical Poissonian
errors related to the total counts present in each radial bin are
particularly small (cf. Fig.~\ref{fig:sbkg}).
They do not consider any fluctuations present in each bin due to
azimuthal variations within each sector. The contribution from these
azimuthal fluctuations will be discussed in the next section.

% so that a proper constraint on
% the significance can be assessed only via a collection of the best-fit
% results on several Monte-Carlo simulations of our data (for example,
% Press et al. 1992, pag. 654ff). Each simulation will be generated from
% the original counts corrected by a random amount proportional to the
% original errors. The corresponding significance is the value quoted in
% the following analysis.

Furthermore, as just noted in previous work (cf. Table~4), there is 
enhanced emission above the central part in all models. 
Thus, we have also analysed the surface brightness profiles avoiding
the central 10 arcmin affected by the presence of the large cooling flow
(as shown in the deprojection analysis, it has a typical $r_{\rm cool}
\sim 0.2$ Mpc, equivalent to about 6.5 arcmin; cf. also the colour
ratios).
As indicated in Table 5, the reduced $\chi^2$ value decreases
significantly but still remains too large to be statistically acceptable.

% A reduced $\chi^2$, median of the values obtained from the best-fit
% between 30 and 80 arcmin, is shown beside the corresponding curve.
% Also the request of reduced $\chi^2$ in the order of unity with a
% hundred of degrees of freedom (as in our fit) indicates a probability
% of about 50 per cent to get a better value for $\chi^2$. 
% This is generally observed in a non-relaxed cluster. 
% Applying our fitting procedure on 1000 simulated surface brightness
% profiles for each of the region analyzed, we observe a significance for
% the original set of the best-fit parameters of: 92; 24; 71; 67 per cent,
% for North, West, South and East, respectively. 
% The significance is improved for all the region, except for West. We
% discuss further this case in the next subsection.

\begin{table*}   \caption{Best fit values of the surface brightness
profiles and the corresponding reduced $\chi^2$, $\chi^2_{\nu}$, adopting
different models. The values quoted are (scale radius -in arcminutes-,
slope). Particularly, we present ($r_{\rm c}, \beta$) for the 
$\beta$-model ($\beta$); ($r_{\rm cut}; r_{\rm P}, \gamma;
r_{\rm c}, \beta$) for the 'power law + $\beta$' model (pow + $\beta$);
($r_{\rm cut}; r_{\rm P1}, \gamma_1; r_{\rm P2}, \gamma_2$) for
a two broken power laws (2 pow); ($r_{\rm s}, \eta$) for the NFW
gas density profile (NFW gas); ($r_{\rm V}, \alpha$) for the de
Vaucouleurs law (de Vauc). 
% the Mellier-Mathez law (MM law); 
} \begin{tabular}{| r c c c |} \hline
Region              & model      &  values  &  $\chi^2_{\nu}$\\ \hline
\multicolumn{4}{c}{\bf only Poissonian error} \\
North: $10'< r <80'$  &  &  ( 8.8, 0.65)  &  1.90 \\
   $0'< r <80'$  & $\beta$  &  ( 3.6, 0.58)  &  6.67 \\
West: $10'< r <80'$  &   &  (12.5, 0.60)  &  12.87 \\
   $0'< r <80'$  & $\beta$  &  ( 2.0, 0.51)  & 51.15 \\
South: $10'< r <80'$  &   &  (25.3, 1.05)  &  2.60 \\
   $0'< r <80'$  & $\beta$  &  ( 2.2, 0.53) & 71.12 \\
East: $10'< r <80'$  &   &  (24.8, 1.23)  &  8.55 \\
  $0'< r <80'$  & $\beta$  &  ( 4.4, 0.60) & 70.89 \\
Centre-off: $10'< r <80'$  &   &  (17.5, 0.85) & 15.42 \\ 
    $0'< r <80'$  &  $\beta$  &  ( 3.4, 0.57)  & 122.38 \\
               &            &     &   \\ 
\multicolumn{4}{c}{\bf Poissonian error and azimuthal fluctuations} \\
North: $10'< r <80'$ & $\beta$ & {\bf (9.5,0.67)} & {\bf 0.12} \\
$0'< r <80'$ & $\beta$ & {\bf (4.3,0.62)} & {\bf 0.17} \\
   & pow + $\beta$  &  {\bf (10.0; 3.5,1.21; 9.5,0.67)} & {\bf 0.14} \\
   & 2 pow & (15.0; 2.0,1.28; 4.3,1.87) & 0.18 \\
   & NFW gas & (15.6,8.87) & 0.22 \\
   & de Vauc & (0.017,0.29) & 0.13 \\
%    & MM law  & (0.6,0.67,0.41) & 0.12 \\
West: $10'< r <80'$ & $\beta$ & {\bf (12.9, 0.61)} & {\bf 0.37} \\
$0'< r <80'$ & $\beta$ & {\bf (2.7, 0.52)} & {\bf 1.00} \\
   & pow + $\beta$  & {\bf (27.0; 3.3,1.38; 67.6,1.67)} & {\bf 0.11} \\
   & 2 pow & (26.0; 2.0,1.38; 4.3,1.77) & 0.34 \\
   & NFW gas & (25.3,7.84) & 1.08 \\
   & de Vauc & (0.001,0.23) & 0.54 \\
%    & MM law  & (0.08,4.4,0.50,0.56) & 0.38 \\
South: $10'< r <80'$ & $\beta$ & {\bf (25.8,1.10)} & {\bf 0.19}\\
$0'< r <80'$ & $\beta$  & {\bf (21.7,0.99)} & {\bf 0.72} \\
   & pow + $\beta$  & {\bf (12.0; 3.4,1.39; 25.6,1.07)} & {\bf 0.23} \\
   & 2 pow & (26.0; 2.0,1.33; 6.1,2.39) & 0.32 \\
   & NFW gas & (60.7,12.79) & 0.66 \\
   & de Vauc & ( 3.33,0.67) & 0.54 \\
%    & MM law  & (0.002,35.6,1.39,1.83) & 0.66 \\
East: $10'< r <80'$ & $\beta$ & {\bf (14.5, 0.88)} & {\bf 0.40} \\
$0'< r <80'$ & $\beta$ & {\bf (10.1, 0.79)} & {\bf 0.64} \\
   & pow + $\beta$  & {\bf (18.0; 3.5,1.22; 6.4,0.79)} & {\bf 0.33} \\
   & 2 pow & (19.0; 2.0,1.22; 5.7,2.30) & 0.33 \\
   & NFW gas & (28.5,10.82) & 0.52 \\ 
   & de Vauc & (0.42,0.46) & 0.50 \\
%    & MM law  & (0.97,0.8,0.04,0.51) & 0.46 \\
Centre-off: $10'< r <80'$ & $\beta$ & {\bf (17.0,0.81)}& {\bf 0.05} \\ 
$0'< r <80'$  & $\beta$ & {\bf (4.7,0.63)} & {\bf 0.54} \\
   & pow + $\beta$  & {\bf (10.0; 3.4,1.24; 17.0,0.81)} & {\bf 0.09} \\
   & 2 pow & (22.0; 2.0,1.31; 5.2,2.12) & 0.14 \\
   & NFW gas & (20.3,9.37) & 0.58 \\ 
   & de Vauc & (0.045,0.33) & 0.13 \\ \hline
%    & MM law  & (0.02,15.6,1.21,0.85) & 0.07 \\
\end{tabular}
\end{table*}

\subsubsection{{\bf Estimated error in the surface brightness mean value}}

\begin{figure*}
\psfig{figure=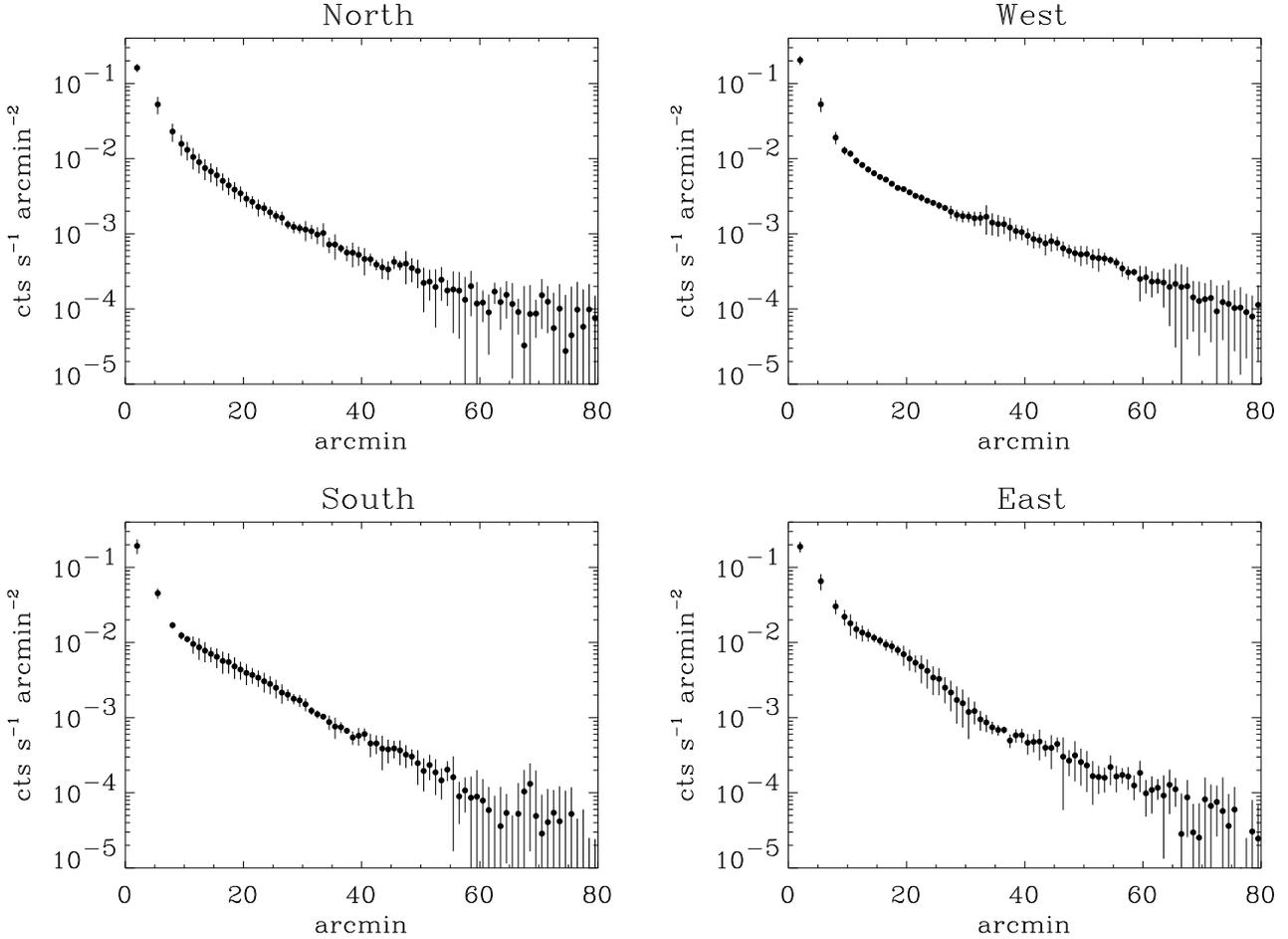,width=\textwidth,angle=90}
\caption{The [ 4, 3, 2, 1, (all the remaining) 1] arcmin plots with the
errors obtained considering also the azimuthal fluctuations within
each radial bin. These plots are fitted with the $\beta$-model and the
other models discussed in section 2.4. The results of the fitting
procedure are presented in Table~5.
} \label{fig:fluc} \end{figure*}

\begin{figure*}
\psfig{figure=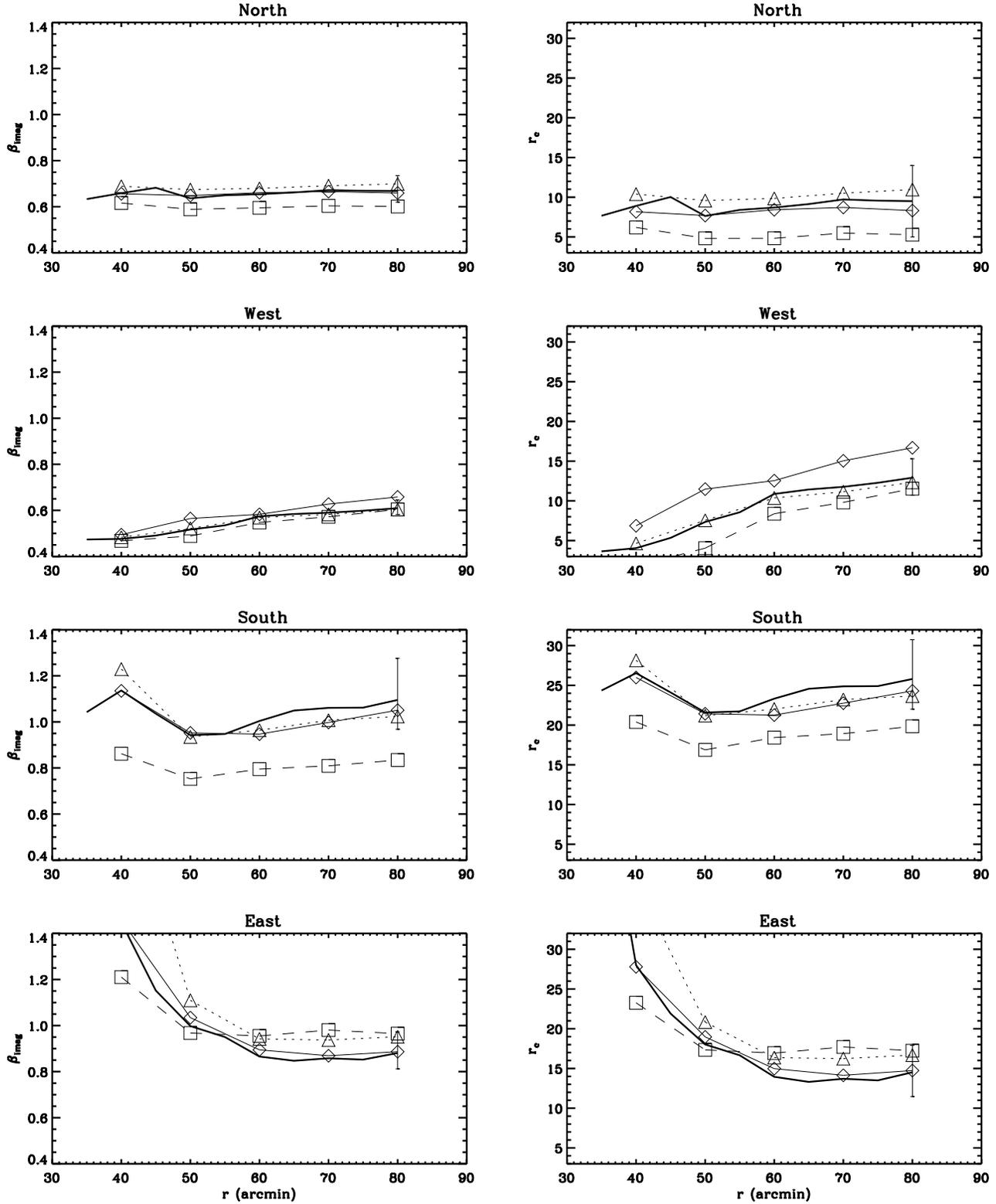,width=\textwidth,angle=0}
\caption{These plots show the results on the $\beta$ value and the core
radius, $r_{\rm c}$, when the $\beta-$model is fitted on the 
profiles binned in [ 4, 3, 2, 1, (all the remaining) 1] arcmin and
selected between 10 arcmin
and different outer radius in the 3 energy bands: soft ({\it triangles}
joined by a dotted line); medium (dashed line and {\it squares}); hard
(solid line and {\it diamonds}). The thickest solid line represents the
result when the $\beta-$model is applied to the 0.5-2.0 keV
exposure-corrected image. At the last point on this line are related the
error-bars drawn. These plots show a correlation between
$\beta$ and $r_{\rm c}$ (see also Jones \& Forman 1984).
} \label{fig:beta} \end{figure*}

\begin{figure}
\psfig{figure=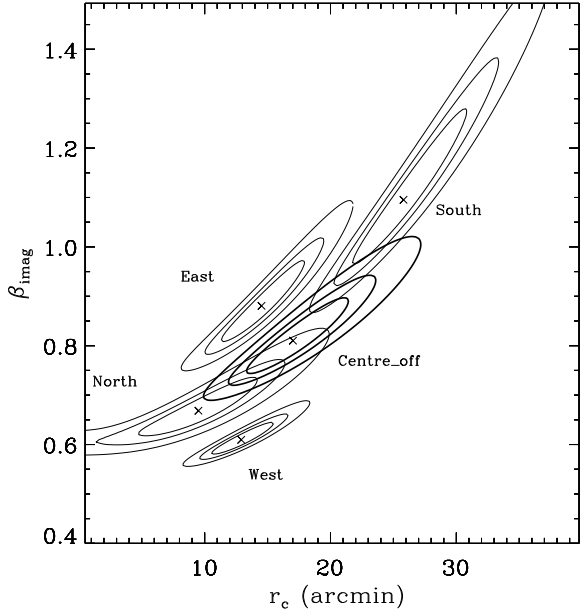,width=0.5\textwidth,angle=0}
\caption{The 68, 90 and 99 per cent confidence contours in the $r_{\rm
c}-\beta$ plane for the $\beta$-model applied on the 0.5--2 keV surface
brightness profile selected between 10 and 80 arcmin 
(cf. the last point on the thickest solid line in Fig.~\ref{fig:beta}).
} \label{fig:cont} \end{figure}

% *** are $\beta$ and colours directly related ? [cf. surbri\_ratio.idl
% and surbri\_theory.idl] ***

\begin{figure}
\psfig{figure=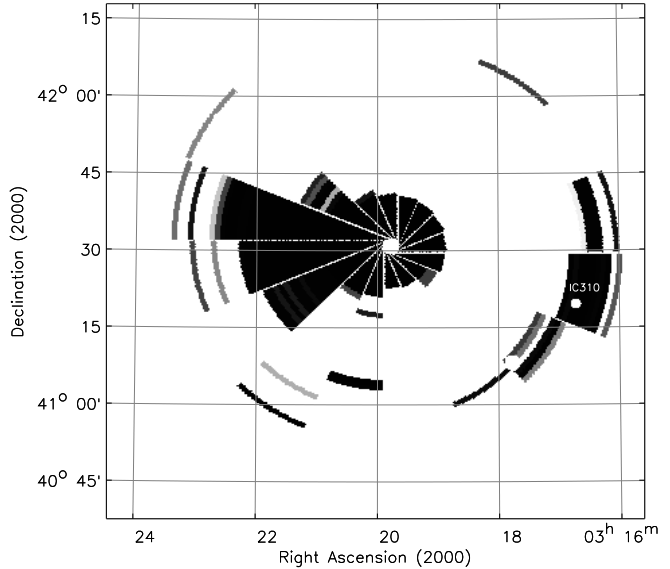,width=0.5\textwidth,angle=0}
\caption{This is a 45\arcmin -radius map of the residuals significant at
the 95 per cent  confidence limit in terms of the $\chi^2$ probability
distribution as drawn by equation~(5). It is obtained after the mosaic of
the four exposure--corrected image has been subtracted from the cluster
emission simulated by a $\beta$--model with
parameters $r_{\rm c}=17.0$\arcmin\ (68 per cent uncertainties range:
13.5, 21.3), $\beta=0.81$ (0.75, 0.90), and $S_0 = 2.31 \times
10^{-2}$ counts s$^{-1}$ arcmin$^{-2}$.  
% normalized to the observed surface brightness between 28 and 30 arcmin. 
With a reduced $\chi^2$ of 0.05, these values provide the best-fit to the
Centre-off surface brightness profile, between 10 and 80 arcmin (cf.
Table~5).
} \label{fig:map} \end{figure}

%  ---------------    TEXT    ----------------

Previously, we have measured the error on the surface brightness as the
Poissonian error related to the total counts in each radial bin. 
This is a statistical error that does not consider any fluctuation on the
mean value (i.e., the surface brightness value) due to azimuthal
variations intrinsic to the cluster.

The contribution expected from these fluctuations is particularly high in
a non-relaxed cluster such as Perseus. We have estimated the contribution from
the dispersion of the mean values (assumed normally distributed) of the surface
brightness calculated in $N_{\rm sect}$ azimuthal sectors. 
This is done in each off-axis image (where the maximum angular size is about
90\degr\ at the maximum width).
Choosing $N_{\rm sect}=4$ (i.e. angular sector of about
15\degr), we obtain a relative error of about 15 per cent over
the radial range [ 0, 80] arcmin, i.e. about 10 times larger than the
simple Poissonian error. 
The surface brightness profiles with the azimuthal fluctuations added in
quadrature to the Poissonian errors are shown in Fig.~\ref{fig:fluc}. 

After fitting these surface brightness profiles, we obtain values of
reduced $\chi^2$ which are low due to the large, but
{\it real}, errors now present in the data (cf. Tab.~4).

Of course, the value of $\chi_{\nu}^2$ is reduced by increasing
$N_{\rm sect}$, as fluctuations further dominate. On the other hand, it
approaches a more statistically meaningful value of about 1 when both the
angular region considered has width $\la$ 15\degr\ and only Poissonian
error is considered. Then, the contribution from fluctuations is still
larger by a factor of 3 and more.

In the following analysis, we quote the best-fit results obtained over
each 90\degr-offset region (cf. Tab.~4), that can be considered as
representative of the variations along different directions.

In Fig.~\ref{fig:beta}, we plot the measured $\beta_{\rm imag}$ and core
radius for each sector, at different outer radii and for each of the three
energy bands. Thus, we check also the dependence of $\beta_{\rm imag}$
upon the colours, that we expect if any merging action is produced by
substructure with a significant difference in temperature from the ICM.

Using the $\beta-$model for the analysis of the different sectors,
we find a value around 0.65 in the North, and an increase from the
exterior to the centre, from 0.45 to 0.65, in the West.

The eastern and southern regions provide values of about 1 over all radial
ranges considered. These values, as shown by Edge \& Stewart (1991),
indicate an on-going merger that seems to involve the eastern region (cf.
again the conclusion in Schwarz et al. 1992) as well as the southern one.
In particular, the soft profile steepens significantly around 40 arcmin
to the East. The colour analysis shows a gradient that
agrees with a softening of the emission (cf. Fig.~\ref{fig:dev4}).

Due to the correlation in the $\beta-r_{\rm c}$ plane (cf.
Fig.~\ref{fig:beta}),
the core radii emphasize a peculiar gas distribution in these 2
sectors: they are about twice the ones estimated in the North and West.
The presence of this correlation can also partially explain the
disagreement between the parameters fitted in the previous work, where 
the lower core radii correspond to lower $\beta$ (cf. Table~4).

Fig.~\ref{fig:beta} also shows that, although there is a trend for
the gas distribution to be flatter in the medium band, the surface
brightness selected in the energy range 0.9-1.3 keV shows a steeper  
profile in the East sector, underlining the ICM interaction with a
clump of gas at temperature greater than 1 keV.

The 0.5-2.0 keV band indicates a $\beta_{\rm imag}$ value consistent with
the average of the 3 other values calculated in the different bands.

% These are slightly consistent one each other. Using a Kolmogorov-Smirnov
% test (ks\_king\_fit.idl), only West shows an agreement among them within
% a 5 per cent confidence level, whereas North is the only one with the
% $\beta$'s from soft--medium--hard profiles not-consistent at the same
% confidence level. 
 
The errors on the two interesting parameters, $\beta_{\rm imag}$ and
$r_{\rm c}$, are shown in Fig.~\ref{fig:cont} as contours plotted at 68,
90 and 99 per cent of confidence level (i.e. $\chi^2$-value regions
of $\chi^2_{\rm min} + 2.30$, $\chi^2_{\rm min} + 4.61$ and
$\chi^2_{\rm min} + 9.21$, respectively; e.g. Press et al. 1992,
page 687). They are estimated with the fit performed on the surface
brightness profile selected between 10 and 80 arcmin in 0.5-2.0 keV band.
No agreement is evident, at the 99 per cent confidence level, among the
different sectors considered. Thus, the complexity of the area in
terms of the gas distribution is once more emphasized. However, a good
representation of the region is provided through the azimuthally averaged
{\it Centre-off} profile. 

With these new constraints on the $\beta_{\rm imag}$ values in the
different sectors of A426, we can address the issue of the
$\beta$-problem.

\subsubsection{{\bf Resolution of the $\beta$ problem}}

Following Bahcall \& Lubin (1994), and the recent results on N-body
simulations of clusters of galaxies (cf. Thomas et al. 1998), we first
consider a correction to the $\beta_{\rm imag}$ value. 
The $\beta$-model, as described by equation~(3), has been introduced
assuming that the gas and the galaxies are in hydrostatic equilibrium with
the cluster potential, and that the galaxy distribution is well
described by the King approximation (1962) to the isothermal sphere. 
This approximation is proportional to $r^{-3}$ at the outer radii, and
does not match the dark matter dependence of $r^{-2.4}$, which is 
determined from N-body simulations (cf. Fig.~\ref{fig:dens}). 
Then, the hypothesis that the {\it optical light} traces the mass, i.e.
that the galaxy number density profile describes the gravitational profile
of the cluster over the virialized region (as the recent CNOC results also
seem to confirm; cf. Carlberg et al. 1997), indicates a way to correct the
$\beta_{\rm imag}$ value to give the correct dependence on the galaxy
distribution of the radius:

\begin{equation}
\rho_{\rm gas} (r) = \rho_{\rm gal}(r)^{\beta_{\rm imag}} \propto r^{-3
\beta_{\rm imag} } \propto r^{-2.4 \beta_{\rm imag, corr} },
\end{equation}
i.e., $\beta_{\rm imag, corr} = 1.25 \beta_{\rm imag}$.

Thus, we have compared $\beta_{\rm spec}=1.09^{+0.23}_{-0.14}$ with the
corrected values of $\beta_{\rm imag}$; 0.84, 0.76, 1.37,
1.10, 1.01 (with 68 per cent errors of about 10 per cent) for the North,
West, South, East and Centre-off, respectively.
There is now a reasonable agreement between $\beta_{\rm imag, corr}$ and
$\beta_{\rm spec}$, which eliminates the $\beta$ problem in the
Perseus cluster.

% Also, we note that by using
% the $\sigma_{\rm dpr}$ value of 850 km s$^{-1}$ (that provides a flat
% temperature profile at 6.3 keV), we calculate a $\beta_{\rm spec}$ of
% about 0.75. 

%%%%%%%%%%%%%%%%%%%%%%%%%%%   DBASE/beta_dpr.idl
Furthermore, assuming the equation $\beta_{\rm spec} = \beta_{\rm imag,
corr}$ is valid, we can calculate a velocity dispersion, $\sigma_{\rm
imag}$, using $\beta_{\rm imag}$ values (from the 0.5-2.0 keV band
selected in the radial range 10--80 arcmin) and an isothermal
temperature of 6.3 keV.
These values are quoted in square brackets in Table~3, and show a good
agreement with the deprojection estimates, $\sigma_{\rm dpr}$.
All the velocity dispersions calculated so, are 
also marginally consistent with the optically determined velocity
dispersion, $\sigma_{\rm opt}$, of 1026$^{+106}_{-64}$ km s$^{-1}$.

Using the best-fit results from the $\beta-$model, we investigate in
Fig.~\ref{fig:map} the residuals (significant at the 95 per cent
confidence level) for a cluster simulated by using a $\beta$ - model,
$S(x,y)$, compared to the raw data, $I(x,y)$. Both $S(x,y)$ and $I(x,y)$
are in counts pixel$^{-1}$. Then, the significance is assessed in terms
of a $\chi^2$ distribution:
\begin{equation}
\chi^2 = \sum^{N_x \times N_y}_{j = 1} \frac{[I(x_j,y_j) - S(x_j,y_j)]^2}
{S(x_j,y_j)},
\end{equation}
where $N_x$ and $N_y$ are the numbers of pixels in the rows and columns in
the selected sectors of 22\fdg5 in annular circular bins of [ 4, 3, 2,
1, (all the remaining) 1] arcmin.

%%%% GROUPS L_X  in  DBASE/lum_group.idl

From this map, apart from the central cooling region which contributes to
the total luminosity $L_{\rm bol} = 2.8 \times 10^{45}$ erg s$^{-1}$
about 50 per cent (cf. the deprojection results in in
Fig.~\ref{fig:dpr}), the eastern excess extends to 40 arcmin from the
centre with an estimated luminosity in the residuals of about $8 \times
10^{43}$ erg s$^{-1}$. From the $L-T$ relationship in the groups and
clusters sample of Mulchaey and Zabludoff (1998), this luminosity
corresponds to about 2 keV, as we also conclude from the colour analysis. 
Other extended emission appears related to the galaxy IC310, at about
40 arcmin westwards (IC310 is the masked point source; see also Cruddace
et al. 1997). This clump has an estimated luminosity of $\sim 1 
\times10^{43}$ erg s$^{-1}$ with a corresponding temperature of 1 keV.
This is further support for the
drastic increase in the colour ratio values for the same region is
attributed to an excess of cool gas, as seen in Fig.~\ref{fig:dev2}. 
Both of these clumps appear as groups still merging with the cluster.

Other poorly defined discrepancies with an azimuthally averaged
$\beta$-model profile appear in the South. 
They can be part of the eastern excess and are responsible of the
$\beta$ value of about unity measured in that sector.

%%%%%%%%%%%%%  king_sim_one.idl  
%%%%%%%%%%%%%  fit on CENTRE_OFF with fluct  by  pers_off.idl
%%%%%%%%%%%%%  COLORS  king_fit.idl,  .5-2  king_fit_one.idl
%%%%%%%%%%%%%  to fit 1 or 2 beta model, w or w/o fluctuations ...
%%%%%%%%%%%%%                          fit_beta-1.0.idl

We stress that these residuals depends on the assumption of the model. As
shown in Section~2, the residuals are not significant if the elliptical
isophotes are allowed to have free centroids. Furthermore, 
modelling this elliptical isophotal intensity profile
with a $\beta$-model in the range [10-80] arcmin, we obtain $\beta = 0.63
\pm 0.01$ and $r_{\rm c} = 9.3 \pm 0.2$ arcmin, for a reduced $\chi^2$ of
33.6. These values agree well with those observed for the sector
to the North, that appears as the most relaxed region in the cluster.

\subsection{On the fitting of the surface brightness profile}

The inadequacy of the $\beta$-model in reproducing the gas profile
over the whole radial range, in particular in the central part,
has driven us to look for other possible functional forms. 

We consider the following models:
\begin{enumerate}
\item a power law plus a $\beta$ model, to describe the inner
cooling region and the outer part, respectively, with a radius of cut-off,
$r_{\rm cut}$, as a free parameter:
\begin{equation}
S_{\rm b} = \left\{ \begin{array}{ll}
S_0 \left(\frac{r}{r_{\rm P}}\right)^{1-2\gamma} & \mbox{if $r<r_{\rm
cut}$} \\
S_0 \left[ 1 + \left(\frac{r}{r_{\rm c}} \right)^2\right]^{0.5-3\beta} &
\mbox{if $r>r_{\rm cut}$}  \end{array}  \right. 
\end{equation}
where the exponent for the 'power law' model comes from the consideration
that 
\begin{equation}
\rho_{\rm gas} = \rho_{0} \left(\frac{r}{r_{\rm P}} \right)^{-\gamma}
\longrightarrow \ S_{\rm b} = S_0 \left(\frac{r}{r_{\rm P}}
\right)^{1-2\gamma} ;
\end{equation}

\item two density power laws, boken at radius $r_{\rm cut}$;

\item a generalized form of the de Vaucouleur's law (1948):
\begin{equation}
S_{\rm b} = S_0 \exp \left[ -\left(\frac{r}{r_{\rm V}} \right)^{\alpha}
\right].
\end{equation}

% \item a generalized form of the Mellier-Mathez law (1987):
% \begin{equation}
% S_{\rm b} = S_0 \left(\frac{r}{r_{\rm MM}} \right)^{-\alpha_1} \exp
% \left[ -\left(\frac{r}{r_{\rm MM}} \right)^{\alpha_2} \right].
% \end{equation}

\end{enumerate}

We also introduce a gas density profile obtained from the NFW dark matter
profile when the hydrostatic equation under the isothermal condition is
applied:
\begin{equation}
\rho_{\rm gas} = \rho_0 (1+x)^{\eta/x},
\end{equation}
where $x= r/r_{\rm s}$ and $\eta = 4\pi G \rho_{\rm s} r^2_{\rm s} \mu
m_{\rm p} / (kT)$ (cf. Makino, Sasaki \& Suto, 1998, and Ettori \& Fabian, 
1998).
The surface brightness profile is then obtained by numerical
integration of the gas density profile in equation~(9).

We use the reduced $\chi^2$ provided by each model to assess their
suitability. The best-fit parameters are listed in Table~5. From this
table, the 'power law + $\beta$ model' appears as the more appropriate
description of the data over all the range of radii.
In terms of the $F$-test, when we compare the $\beta$ model with the
'power law + $\beta$' fit, we obtain that it is still a good
description of the northern data, but it becomes worst at 99 per cent of
confidence level in the cases of the West, South, East
and Centre-off.
%%%%%%%%%%%%%%%%%%%%%%%  f_stat.idl 
Once again (cf. the colour analysis and Fig.~18), the North appears as the
more relaxed region in the Perseus cluster.

We also note that the 'power law + $\beta$ model' highlights the merger
of the clump of gas around IC310 in the outer region to West. In fact, the
$\beta$ value increases there to about 1.4, compared to a value of 0.6
from the single $\beta$ model. Moreover, the cut-off radius is 26 arcmin
to West and 18 arcmin to East (around 12 in the other sectors), 
underlining the elliptical configuration discussed in Section~2.

However, the $\beta$ model is a poor description when it is also compared
to the  models (ii) and (iii), but it is better than the gas profile
obtained from the NFW law. In fact, this
latter profile shows the greatest disagreement with the data. We check
elsewhere on the consistency of the gas distribution obtained from NFW
dark matter potential with a larger set of X-ray cluster profiles.

The slope in $\rho_{\rm gas} \propto r^{-\alpha}$ is in the range
[1.22, 1.38] in the core region and become steeper to [1.77, 2.39] in the
outer part of the Perseus cluster. There, the slopes from the two
broken power laws are consistent, within an average error of 15 per cent,
with the asymptotic behaviour of the $\beta$ model (apart from the
southern region where the discrepancy is 28 per cent).

\section{The constraints on the mass profile}

\begin{figure*}
\hbox{ \psfig{figure=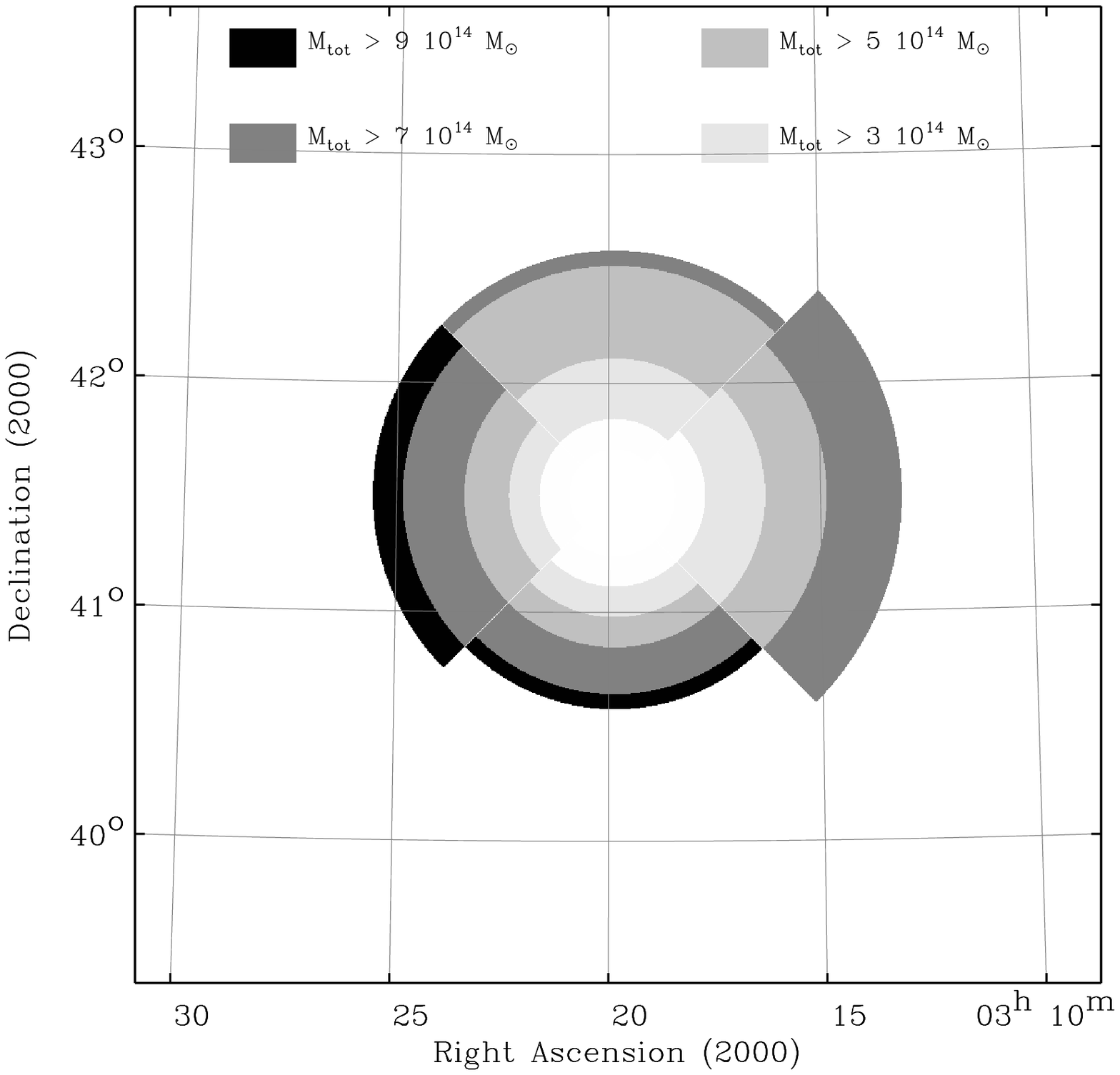,width=0.5\textwidth,angle=0}
\psfig{figure=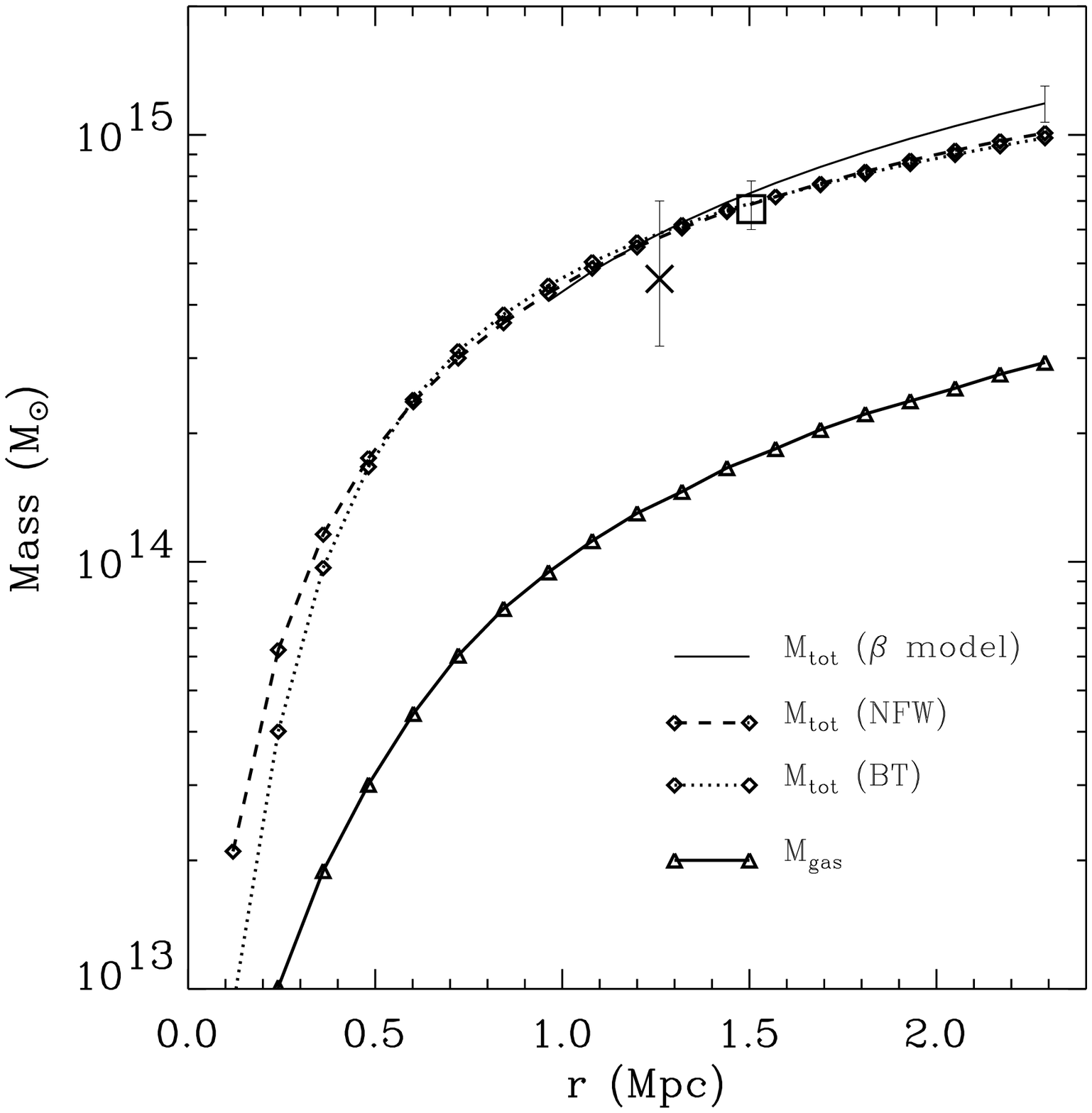,width=0.5\textwidth,angle=0} }
\caption{These plots present the mass distribution how has been 
obtained applying the deprojection technique, under the hydrostatic and
isothermal assumptions. (left) The map is a 
combination of the four off-axis observations results on the total 
gravitating mass. (right) The mass profiles obtained by applying
both the deprojection technique and the best-fit results from the
$\beta$-model on the azimuthally averaged surface brightness profiles
% (both {\it Centre-on} and {\it Centre-off}, to 1.1 and 1.8 Mpc,
% respectively) 
are here compared with the previous
estimates: about $5 \times 10^{14} M_{\odot}$ at R=1.3 Mpc ({\it cross};
Eyles et al. 1991); $6.7 \times 10^{14} M_{\odot}$ at 1.5 Mpc
({\it square}; from Cruddace et al. 1997 after we adopt the redshift of
0.0178). The last point at 2.3 Mpc reports a value of
$1.19^{+0.12}_{-0.12} \times 10^{15} M_{\odot}$, for the range of the 10th
and 90th percentile on 100 Monte-Carlo simulated estimations (see Sect.~4).
} \label{fig:mass} \end{figure*}

\begin{figure}
\psfig{figure=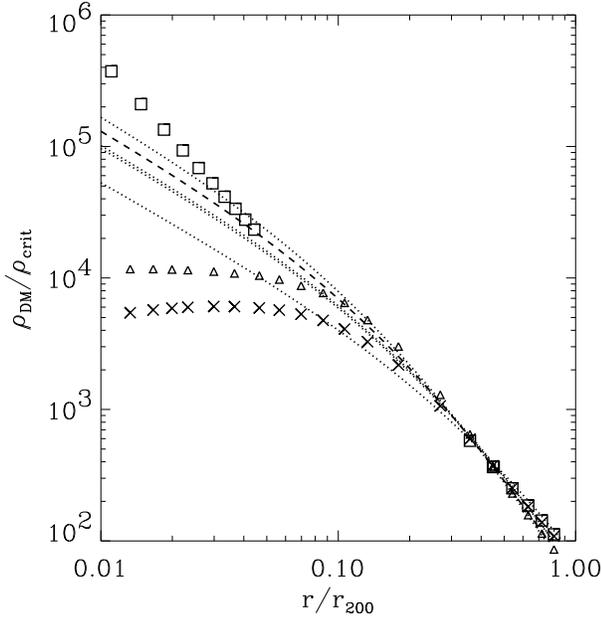,width=.5\textwidth,angle=0}
\caption{The dark matter density profiles as obtained from the best-fit
values of the NFW gas profile are the dashed line ({\it Centre-off}) and
the dotted lines (the other four sectors). The {\it crosses} represent the
result from the $\beta$ model, the {\it triangles} are from the
deprojection analysis, and the {\it squares} what we obtain applying 
two broken power laws (we plot just the external values for a more
clarity). All of them are obtained in the case of {\it Centre-off}.
} \label{fig:slop} \end{figure}

\begin{figure*}
\psfig{figure=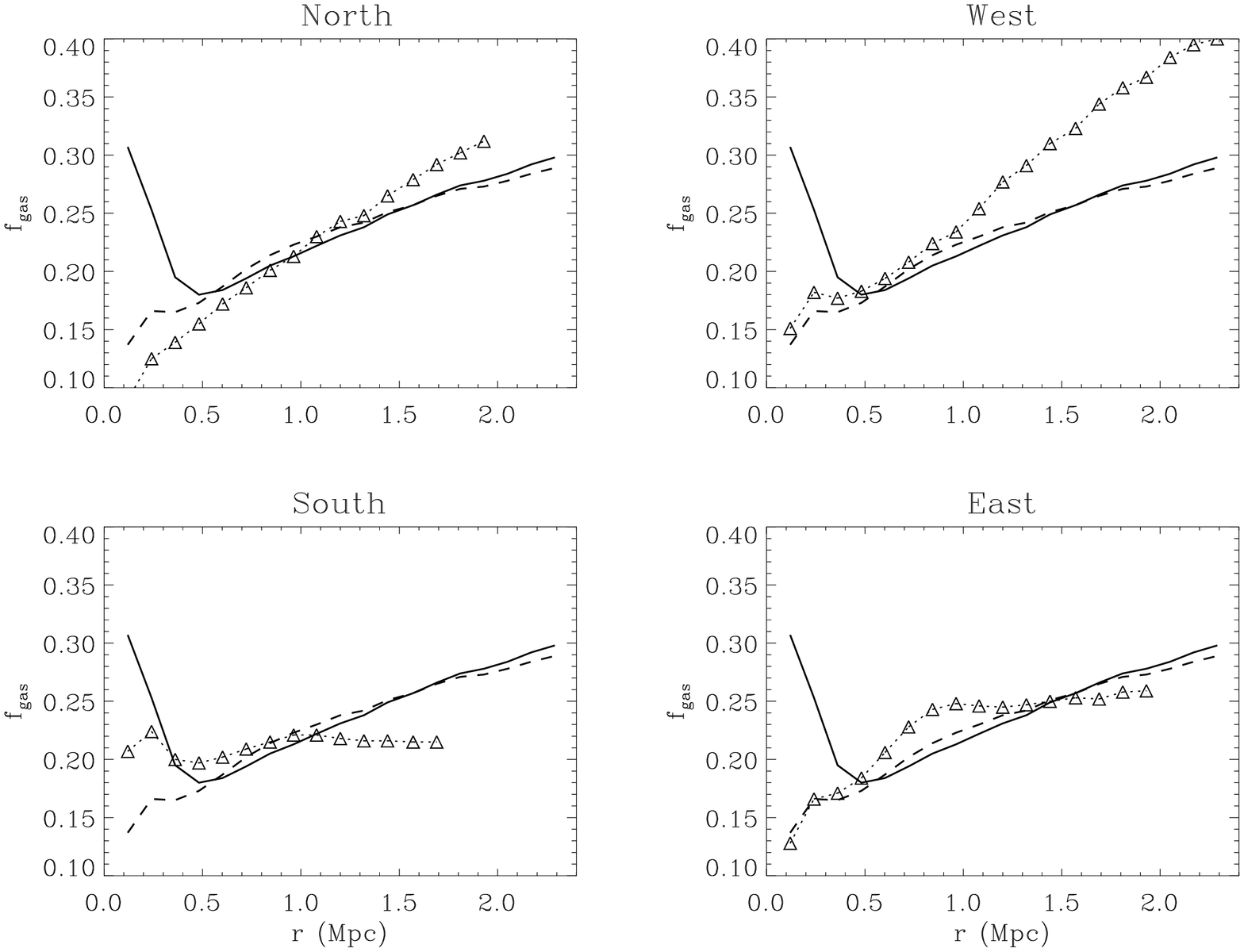,width=\textwidth,angle=90}
\caption{These plots show the gas fraction, $f_{\rm gas}$, as function of
the radius $R$ for the four sectors deprojected by assuming a
Navarro-Frenk-White profile dark matter density profile (triangles
joined by a dotted line). 
%% In each panel, we draw the on-axis and off-axis deprojection result. 
We also show, as comparison, the value for the gas fraction measured in
the azimuthally averaged deprojected image, both using a BT profile 
%% (the two solid lines, for the on-axis image and the combination of the
%% off-axis observations) 
(solid line) and a NFW profile under isothermal assumption (dashed line).
(Cf. also Table~3 for the values of $f_{\rm gas}$ quoted at $R_{\rm out}$
and the relative error).
At 1.5 Mpc, Cruddace et al. (1997) calculate a value of about 0.24.
} \label{fig:fgas} \end{figure*}

The form of the gravitational potential remains the most uncertain
element of the deprojection analysis, since the observational constraints
are not completely sufficient. 
However, some conclusions on the shape of the dark matter profile
can be achieved under the assumptions of hydrostatic equilibrium and
isothermality of the gas distribution in the cluster. Then,
the total gravitating mass, responsible for the external potential that
maintains the gas mass within a radius $R$ in equilibrium, is written as
\begin{equation}
M_{\rm tot} = \int^r_0 4\pi \ \rho_{\rm DM} \ r^2 dr =
-\frac{r^2}{G \rho_{\rm gas}} \frac{dP_{\rm gas}}{dr}.
\end{equation}

This equation, together with the perfect gas law, is used in
Fig.~\ref{fig:mass} to describe how the total gravitating mass depends on
the radius, applying both the deprojection technique and the
$\beta$-model. 
The results indicate consistency between the total mass, estimated by
these methods, and previous estimates (when the errors are properly
considered).
In the same figure, the mass distribution, as obtained through the
deprojection analysis, emphasizes the over-density in the east sector where
the merging group acts.

In Fig.~\ref{fig:slop}, we plot the dark matter density profile as
obtained from (i) the best-fit ($\eta, r_{\rm s}$) values of NFW gas
profile, once the definition of $\eta$ is used with a plasma temperature
of 6.3 keV; (ii) the best-fit slopes $\gamma$ from the two broken power
laws model, given that $M_{\rm tot} \propto \gamma \ kT \ r$ and
$\rho_{\rm DM} \propto \gamma \ kT \ r^{-2}$; (iii) the $\beta-$model,
in which $M_{\rm tot} \propto \beta \ kT \ r^3/(r^2 + r_{\rm c}^2)$ and
$\rho_{\rm DM} \propto \beta \ kT \ (r^2 + 3r_{\rm c}^2)/(r^2 + r_{\rm
c}^2)^2$; (iv) the result of the deprojection analysis using the potential
from the true isothermal sphere. 
%% and the ratio $\lambda = \rho_{\rm gas,2}/\rho_{\rm gas,1}$, $M_{\rm
%% tot} = ( M_{\rm tot, 1} + \lambda M_{\rm tot, 2}) /(1+\lambda)$.
Here we note that both the $\beta$ and the power law models indicate
a dark matter density profile that is flatter than N-body simulations
profile, with a dependency upon radius of slope 2, instead of
about 2.4. However, on this radial range, this affects slightly the
determination of the total mass as shown in Fig.~\ref{fig:mass}.

From fitting the core with a power law, it appears more convenient to
use a steeper profile than the NFW one, whereas the King profile provides
a plateau (cf. also Fig.~\ref{fig:dens}). 
Furthermore, the best-fit values define $r_{200}$, the radius at which the
average over-density in the cluster is 200, equal to about 2.7 Mpc.
This means that we are mapping, in our analysis, the region where the
over-density is about 300 or more.

The results in Fig.~\ref{fig:slop} show good agreement within the region 
0.4 -- 0.8 $r_{200}$, i.e. above the core region that depends upon the
functional form of the potential adopted and within the radial limits of
our observations. 
% In that range, the dark matter density slope from the
% different methods (i)-(iv) are $2.43 \pm 0.11, 2.00 \pm 0.02, 1.93 \pm
% 0.42$ and $2.45 \pm 0.06$, respectively (the errors are 1 standard
% deviation on 100 Monte-Carlo simulated dark matter profiles, once the 1
% $\sigma$ errors of the best-fit parameters for the {\it Centre-off}
% profile are defined from the standard deviation of the best-fit values
% obtained in the 4 sectors).

These constraints on the total mass are applied to describe the
distribution of the gas fraction, $f_{\rm gas}$, shown in
Fig.~\ref{fig:fgas}, in dependence both of the central azimuthally
averaged observations and of each sector considered. 
We obtain a gas contribution to the total mass of 30 per
cent at 2.3 Mpc. This value is consistent with the {\it Spartan 1}
estimate at 1.5 Mpc of about 24 per cent (from Fig.~3 in Cruddace
et al. 1997).

Here we note that if any decrease in the average temperature of the
cluster plasma is detected in the outskirts, as appears common in
clusters from the recent analyses of {\it ASCA} data by Markevitch et al.
(1997), then the gas fraction obtained assuming an isothermal
profile should be increased significantly due to the corresponding
reduction of gravitating mass. However, from the colour
analysis, we observe that the gas temperature in Perseus does not decrease
\footnote[4]{using the same analysis, it is difficult to put any upper
limit; cf. Fig.~\ref{fig:meka}} with respect to the reference value of 6.3
keV up to 50 arcmin (about 10 $r_{\rm c}$), at least
in the North and South. In the other two sectors, we detect cool
emission from extended and identified components, probably due to a
merger that could affect the hydrostatic assumption.

The latter two excesses in emission are also seen in the plots of the gas
fraction (Fig.~\ref{fig:fgas}), where the distribution of each sector is
compared to the azimuthally averaged value.
The eastern excess, between 0.5 and 1 Mpc, corresponds to the merging
group, while the excess in the West, where the gas fraction reaches a
value of about 40 per cent at 2.3 Mpc, shows a discrepancy with the
azimuthally averaged value over all the radial range outside 1 Mpc. 
This is also shown in Fig.~\ref{fig:map}, where the
contribution from the extended emission around IC310 becomes significant.

% Allowing the ICM temperature to vary, we require larger core
% radius (of about 1 Mpc) and $kT$ of about 7 keV to match the deprojected
% temperature profile. It is marginally consistent
% with the {\it GINGA} result, and agrees with the higher temperature
% constraints obtained from some other missions, like OSO--8 (Serlemitsos
% et al. 1977) and {\it Tenma} (Okumura et al. 1989). The total mass is
% then larger by 70 per cent with respect to the value estimated in {\it
% Centre-off}.

\section{CONCLUSION}

Our main conclusions are as follows:

\begin{enumerate}

\item  The distribution of the ICM in the Perseus cluster is complex,
being more homogeneous in the northern region, with
clear excess in the eastern part. We can not confirm any temperature
higher than about 6 keV in the North, or North-West areas, as claimed from
the {\it ASCA} results. 

The analysis of the PSPC colour ratios indicates an isothermal profile
around 6 keV within 50 arcmin, a significant decrease in the cooling
flow region, and fluctuations both in the eastern (between 20 and
50 arcmin from the X-ray center) and in the western areas (around IC310,
35--50 arcmin from the center), where we detect cool extended emission
with luminosity of few times $10^{43}$ erg s$^{-1}$. This result
is also supported from a $\beta$-model analysis and shows the presence of 
groups still merging with the main body of the cluster. 
This merger event produces a disturbance of the gas within the
gravitational potential, that can be modelled by elliptical isophotes with
free centroids. This model then eliminates significant residuals 
from the regions of the clumps, emphasizing that these features must 
rather be entities within the cluster, and not foreground / background 
objects.
% However, it should be borne in mind that these 'clumps' are probably
% not physically distinct entities within the cluster (or even foreground
% /background objects) as isophotes fits with free-floating centroids, results
% in a model with no significant residuals in the regions of the clumps.
% Thus, the apparent clumps are really a manifestation of isophote shifts
% caused by a disturbance of the gas within the gravitational potential,
% probably through a merger event.

Furthermore, we show that the observed discrepancy in surface brightness 
can not be justified simply in the basis of the gradient in absorption
present over the region, even through some observed gradient is present as
detected from our colour analysis.

%  \item the Navarro-Frenk-White dark matter profile is more steep than 
%  the one from a true isothermal sphere. Then, we require larger scale
%  radius and velocity dispersion to match the deprojected results
%  provided by the potential from a true isothermal sphere.

\item  We use the $\beta$-model to highlight the disturbed intracluster 
gas distribution in the East, South-East.
The best-fit parameter $\beta_{\rm imag}$ is shown to be dependent on the
outer radius in the extraction of the surface brightness profile, although
it converges outside 50 arcmin (1.5 Mpc).
The disagreement between $\beta_{\rm spec}$ and $\beta_{\rm
imag}$ disappears when $\beta_{\rm spec}$ is properly evaluated 
with a recent estimate of the optical velocity dispersion that takes
into account substructures in the velocity space, and $\beta_{\rm imag}$
is carefully estimated in the outer part of the cluster with a proper 
correction for the slope of the dark matter density profile.

Thus, we can claim that the $\beta$-problem in the Perseus cluster is
solved.

\item The surface brightness profile in the Perseus cluster does not seem
to be well described by a single $\beta$-model over all the range [ 0, 80]
arcmin. This is common for cooling flow clusters that also show merging
in the outskirts, as in Perseus.

After testing several different models, we find that the best statistical
agreement between the data and the gas distribution is obtained with a
model with a power law fitted to the inner cooling region, and a
$\beta$ model for the outer component.

We also apply of the gas profile obtained using the NFW dark matter
potential in the hydrostatic equation. 
However, this does {\it not} provide an acceptable description of the
Perseus data, due to the no-relaxed state of the cluster.

\item With the assumption of hydrostatic equilibrium, we place constraints
on the non-luminous mass density profile from the best-fit results of the
NFW gas profile, $\beta$-model, two broken power laws model, and
the conditions provided from the deprojection analysis.

All of these are consistent in defining $r_{200}$ at 2.7 Mpc. 

The total mass in cluster within 2.3 Mpc (0.85 $r_{200}$)
is 1.19$^{+0.12}_{-0.12} \times 10^{15} M_{\odot}$, where the errors are
the 10th and 90th percentile calculated estimating 100 times the mass
through the $\beta-$model, once the temperature is randomly selected
around its best value and the errors are considered, the core radius is
also randomly selected and $\beta$ is fixed from the best fit on the
surface brightness profile.
The different models agree (within the respective error)
on this estimate, even if a slight difference ($\sim 20$ per cent) is
present in the slope of their dark matter profiles in the range of
0.4--0.8 $r_{200}$, where our observations of the gas profile are
available and not affected from the core region.

\item Using constraints on the gas and non-luminous mass
distribution, we measure the gas fraction to be equal
to about 30 per cent at 2.3 Mpc. This is similar to that observed in Coma
and other clusters (White et al. 1993, White \& Fabian 1995), and
disagrees with the primordial nucleosynthesis estimates of the baryon
fraction in the Universe (e.g. Olive 1996), unless the ratio
between the total density and the critical value, $\Omega_{0, \rm 
matter}$, is less than 0.2.

\end{enumerate}

\section*{ACKNOWLEDGEMENTS} We are grateful to the members of the IoA
X-ray Group for useful discussions.
SE acknowledges support from PPARC and the Cambridge European 
Trust, ACF the support of the Royal Society and DAW that of PPARC.
This research has made use of data obtained through the High Energy 
Astrophysics Science Archive Research Center Online Service, provided 
by the NASA-Goddard Space Flight Centre.


\begin{thebibliography}{}
\bibitem[]{} Allen, S.W., Fabian, A.C., Johnstone, R.M., Nulsen, P.E.J.
\& Edge, A.C., 1992, MNRAS, 254, 51
\bibitem[]{} Allen, S.W. \& Fabian, A.C., 1994, MNRAS, 269, 409 
\bibitem[]{} Allen, S.W. \& Fabian, A.C., 1997, MNRAS, 286, 583
\bibitem[]{} Andreon, S., 1994, A\&A, 284, 801
\bibitem[]{} Arnaud, K.A. et al., 1994, ApJ, 436, L67
\bibitem[]{} Arnaud, K.A., 1996, Astronomical Data Analysis Software and
Systems V, eds. Jacoby G. and Barnes J., p. 17, ASP Conf. Series vol. 101
\bibitem[]{} Bahcall, N.A. \& Lubin, L.M., 1994, ApJ, 426, 513
\bibitem[]{} Binney, J. \& Tremaine, S., 1987, {\it Galactic Dynamics},
Princeton University Press, p. 226f
% \bibitem[]{} B\"ohringer, H., Voges, W., Fabian, A.C., Edge, A.C. \&
%  Neumann, D.M., 1993, MNRAS, 264, L25
\bibitem[]{} Branduardi-Raymont, G., Fabricant, D., Feigelson, E.,
Gorenstein, P., Grindlay, J., Soltan, A. \& Zamorani, G., 1981, ApJ, 248,
55 
\bibitem[]{} Buote, D.A. \& Canizares, C.R., 1996, ApJ, 457, 565
\bibitem[]{} Carlberg, R.G. et al., 1997, ApJL, 485, L13
\bibitem[]{} Cavaliere, A. \& Fusco-Femiano, R., 1976, A\&A, 49, 137
\bibitem[]{} Cruddace, R.G., Kowalski, M.P., Fritz, G.G., Snyder, W.A., 
Fenimore, E.E. \& Ulmer, M.P., 1997, ApJ, 476, 479
\bibitem[]{} de Vaucouleurs, 1948, Ann. Astrophys., 11, 247
% 1958, in {\it Handbuch der Physik}, vol. 53,
% p. 311, ed. S. Fl\"ugge (Berlin: Springer-Verlag)
\bibitem[]{} Dickey, J.M. \& Lockman, F.J., 1990, ARAA, 28, 215
\bibitem[]{} Edge, A.C. \& Stewart, G.C., 1991, MNRAS, 252, 428
\bibitem[]{} Ettori, S. \& Fabian, A.C., 1998, MNRAS, submitted
\bibitem[]{} Eyles, C.J., Watt, M.P., Bertram, D., Church, M.J., Ponman, 
T.J., Skinner, G.K. \& Willmore, A.P., 1991, ApJ, 376, 23
\bibitem[]{} Fabian, A.C., Hu, E.M., Cowie, L.L. \& Grindlay, J., 1981, 
ApJ, 248, 47
\bibitem[]{} Fadda, D., Girardi, M., Giuricin, G., Mardirossian, F. \&
Mezzetti, M., 1996, ApJ, 473, 670
\bibitem[]{} Forman, W., Kellogg, E., Gursky, H., Tananbaum, H. \&
Giacconi, R., 1972, ApJ, 178, 309
\bibitem[]{} Gursky, H., Kellogg, E., Leong, C., Tananbaum, H. \&
Giacconi, R., 1971, ApJL, 165, L43
\bibitem[]{} Hasinger, G., Boese, G., Predehl, P., Turner, T.J., Yusaf,
R., George, I.M. \& Rohrbach, G., 1993, {\it MPE/OGIP Calibration Memo
CAL/ROS/93-015} 
\bibitem[]{} Jones, C. \& Forman, W., 1984, ApJ, 276, 38
\bibitem[]{} Kaastra, J.S., 1992, {\it An X-Ray Spectral Code for
Optically Thin Plasmas} (Internal SRON-Leiden Report, updated version 2.0)
\bibitem[]{} Kent, S.M. \& Sargent, W.L.W., 1983, AJ, 88, 697
\bibitem[]{} King, I. R., 1962, AJ, 67, 471
\bibitem[]{} Kowalski, M.P., Cruddace, R.G., Snyder, W.A. \& Ulmer, 
M.P., 1993, ApJ, 412, 489
\bibitem[]{} Liedahl, D.A., Osterheld, A.L. \& Goldstein, W.H., 1995, ApJ,
438, L115
\bibitem[]{} Makino, N., Sasaki, S. \& Suto, Y., 1998, ApJ, 497, 555
\bibitem[]{} Malumuth, E. M. \& Kirshner, R.P., 1985, ApJ, 291, 8
\bibitem[]{} Markevitch, M., Forman, W.R., Sarazin, C.L. \& Vikhlinin, A.,
1997, ApJ, submitted (astro-ph/9711289)
% \bibitem[]{} Mellier, Y. \& Mathez, G., 1987, A\&A, 175, 1
\bibitem[]{} Mohr, J.J., Fabricant, D.G. \& Geller, M.J., 1993, ApJ, 413,
492
\bibitem[]{} Morrison, R. \& McCammon, D.M., 1983, ApJ, 270, 119
\bibitem[]{} Mulchaey, J.S. \& Zabludoff, A.I., 1998, ApJ, 496, 73
\bibitem[]{} Navarro, J.F., Frenk, C.S. \& White, S.D.M., 1995, MNRAS,
275, 720
% \bibitem[]{} Okumura, Y., Tsunemi, H., Yamashita, K., Matsuoka, M.,
% Kayama, K., Hayakawa, S., Masai, K. \& Hughes, J.P., 1988, PASJ, 40, 639
\bibitem[]{} Olive, K.A., 1996, astro-ph/9609071
\bibitem[]{} Peres, C.B., Fabian, A.C., Edge, A.C., Allen, S.W.,
Johnstone, R.M. \& White, D.A., 1998, MNRAS, in press
\bibitem[]{} Press, W.H., Teukolsky, S.A., Vetterling, W.T. \& Flannery,
B.P., 1992, {\it Numerical Recipes}, Cambridge University Press
\bibitem[]{} Primini, F., et al., 1981, ApJL, 243, L13
\bibitem[]{} Raymond, J.C. \& Smith, B.W., 1977, ApJS, 35, 419
\bibitem[]{} Roettiger, K., Loken, C. \& Burns, J.O., 1997, ApJSS, 109,
307
\bibitem[]{} Rothschild, R.E., Baity, W.A., Marscher, A.P. \& Wheaton,
W.A., 1981, ApJL, 243, L9
\bibitem[]{} Schindler, S. \& M\"uller, E., 1993, A\&A, 272, 137 
\bibitem[]{} Schombert, J.M., 1987, ApJS, 64, 643
\bibitem[]{} Schwarz, R.A., Edge, A.C., Voges, W., B\"ohringer, H., 
Ebeling, H. \& Briel, U.G., 1992, A\&A, 256, L11
% \bibitem[]{} Serlemitsos, P.J., Smith, B.W., Boldt, E.A., Holt, S.S. \&
% Swank, J.H., 1977, ApJL, 211, L63
\bibitem[]{} Slezak, E., Durret, F. \& Gerbal, D., 1994, AJ, 108, 1996
\bibitem[]{} Snowden, S.L., McCammon, D., Burrows, D.N. \& Mendenhall,
J.A., 1994, ApJ, 424, 714
\bibitem[]{} Snyder, W.A., Kowalski, M.P., Cruddace, R.G., Fritz, G.G., 
Middleditch, J., Fenimore, E.E., Ulmer, M.P. \& Majewski, S.R., 1990, ApJ,
365, 460
\bibitem[]{} Thomas, P.A. et al., 1998, MNRAS, 296, 1061
\bibitem[]{} White, D.A., Fabian, A.C., Allen, S.W., Edge, A.C., Crawford,
C.S., Johnstone, R.M., Stewart, G.C. \& Voges, W., 1994, MNRAS, 269, 589
\bibitem[]{} White, D.A. \& Fabian, A.C., 1995, MNRAS, 273, 72 
\bibitem[]{} White, D.A., Forman, W. \& Jones, C., 1997, MNRAS, 292, 419
\bibitem[]{} White, S.D.M., Navarro, J.F., Evrard, A.E. \& Frenk, C.S., 
1993, Nature, 366, 429
% \bibitem[]{} Zabludoff, A.I., Huchra, J.P. \& Geller, M.J., 1990,
% ApJS, 74, 1
\end{thebibliography}
\end{document}